# Direct Numerical Simulation of Fully Developed Turbulent Channel Flow Based on the Corrected Navier-Stokes Equations

Yifan Wang[1], Duo Wang[1], Shuaichen Zhu[1] and Hongyi Xu[1,*]

[1]*Department of Aeronautics and Astronautics, Fudan University, Shanghai 200433, China*

*Corresponding author: hongyi_xu@fudan.edu.cn

**Abstract:** Direct numerical simulations (DNS) of fully developed turbulent channel flows at $Re_\tau = 550$ were performed to investigate the corrected Navier-Stokes (CNS) equations. Grounded in the fluid kinematics of Rortex, the CNS abandons the Stokes' isotropic hypothesis and applies the shearing-only constitutive relation by explicitly eliminating the controversial stretching terms in the stress tensor. Comparisons with the DNS data from the traditional Navier-Stokes (TNS) suggest that the CNS inherently rectifies the near-wall momentum transport. The removal of stretching-induced dissipation shifts the inner- and buffer-layer boundaries towards the wall and effectively suppresses the overshoot in the mean velocity profile in TNS. The turbulence statistics demonstrate a multiscale kinetic energy redistribution that intensifies the near-wall production-dissipation cycle. Furthermore, the topological delineation of instantaneous coherent structures, namely the discovery and definition of rotational/non-rotational interface (RNRI) via the velocity gradient tensor (VGT) discriminant ($\Delta = 0$), confirms that the CNS is capable of capturing the highly complex and interwoven vortical structures. Ultimately, the spectral proper orthogonal decomposition (SPOD) unveils and elucidates that the shearing-only mechanisms intrinsically modulate the spatiotemporal energy cascade, promoting denser and more inclined vortices while enhancing turbulence intermittency by fragmenting the coherent packets. Overall, by isolating and detecting the shearing-only mechanism with physics purity, the DNS based on the CNS provides a more refined perception of the intrinsic dynamics of wall-bounded turbulence, offering a more physical soundness model to capture the interactions among the multiscale coherent structures and the improved capability in predicting the wall-bounded turbulence.

**Key words:** Direct numerical simulation, Channel flow, Navier-Stokes equations, Spectral proper orthogonal decomposition

## 1 Introduction

The Navier-Stokes equations (N-S Eq.; Navier 1827; Stokes 1845) have been commonly accepted by the fluid mechanics community as the dynamics governing equations for quite a number of years with the validity and applicable scopes being well established by the research of many distinguished scholars, which spanned from those earlier days, such as the works of Prandtl (1905), Taylor (1935), Reynolds (1895), von Kármán (1930), and Kolmogorov (1941), to the modern time of the supercomputing era using Reynolds-averaged Navier-Stokes (RANS; Jones & Launder 1972), Large Eddy Simulation (LES; Smagorinsky 1963) and up to the present days applying Direct Numerical Simulation (DNS; Kim et al. 1987) approaches to a variety of flow scenarios. Over the wide time span, the equations have been demonstrated to possess the generally acceptable capability to capture and predict the complex fluid dynamic phenomena, such as laminar flows, boundary layers, separations, transitions and even turbulence. However, due to the recent progress in the vortex identification (VI) research initiated by Liu et al. (2018), the fluid kinematics was essentially revamped through finding the third-generation of VI quantity (3rd-G VI), namely the Rortex/Liutex (hereinafter referred to as Rortex) in the systematical forms of scalar, vector and tensor, which, from the fluid science perspective, provided not only the rationale but also the necessity to carefully revisit the foundations laid out by those fluid mechanics pioneers, such as Helmholtz, Euler, Navier, Stokes, Cauchy and even back to

Newton.

From the fluid kinematics perspective, the velocity gradient tensor (VGT) has always been deemed the key quantity for studying the deformation of a fluid element. The pioneering research was conducted by Cauchy (1828) and Stokes (1845), which was historically well-known as the Cauchy-Stokes (C-S) decomposition, resulting in the symmetric strain rate tensor and the anti-symmetric vorticity tensor being regarded as the fundamentals delineating fluid deformation. Later on, Helmholtz (1858) applied the curl operator to the velocity vector field and defined the quantity as the vorticity vector with its three components coinciding with the three nonzero elements in the vorticity tensor. Thereafter, the vorticity vector was considered as the physical quantity for representing the quite popular and distinctive flow phenomenon of vortex, namely the first-generation of vortex identification (1st-G VI) quantity.

The C-S decomposition can physically be understood as the kinematic preparation to pave the way into fluid dynamics by distinguishing the symmetrical and antisymmetric portions in a VGT for applying to the constitutive relation. However, the recent emergence of Rortex quantities or the 3rd-G VI campaign (Liu et al. 2018) suggests that the vorticity tensor has not reached to its fundamental physics-purity level and the tensor needs to be further decomposed based on the VGT eigenvector, resulting in the discovery of Rortex rigid-rotating tensor and anti-symmetric shearing tensor with their clean physical interpretation, namely the local rigid-rotation axis being in line with the eigenvector direction and the local magnitude representing the local angular rotation speed. The same logic was then applied to the symmetric strain-rate tensor, which was further decomposed into the off-diagonal components of pure shearing and the diagonal components of stretching under the eigenvector directions (Zhu et al. 2023) for the purpose of revisiting the fluid dynamics after the kinematics being renovated by Rortex.

Given the fluid kinematics being clearly understood, the establishment of fluid dynamic equations then needs to be focused on the symmetric strain-rate, which is linearly correlated with the stress tensor by the viscosity coefficient ($\mu$) in the traditional constitutive relation. The N-S Eq. can thereby be derived through substituting the constitutive relation into the momentum conservation equations. Within the context, the traditional constitutive relation of Newtonian fluids, such as water or air, has to be scrutinized to clarify the correlation mechanisms between each component in the stress tensor and the fundamental deformations of pure shearing and stretching, given the fact that the fluid kinematics was essentially revamped by Rortex.

The physical foundation of Newtonian constitutive relation was historically grounded in the Newton's fluid friction law (Newton 1687) and was subsequently confirmed by precisely determining the viscosity coefficient ($\mu$) in a variety of measuring devices, such as the capillary viscometer (Poiseuille 1846), the falling-sphere device (Stokes 1851), and the rotating cylinder viscometer (Couette 1890). The Newton's frictional law was then well established with the mathematical expression of $\tau_{xy} = \mu \partial u / \partial y$. The law manifests the fact that viscous fluid always generates a shearing force when subjected to shearing deformation, which evoked the campaign promoting the fluid dynamics from the Euler equations (Euler 1757) for ideal or inviscid fluid to the N-S Eq. for the viscous fluids (Navier 1827; Stokes 1845). However, when generalizing the Newton's frictional law into a three-dimensional (3D) scenario, Stokes had to resort to the isotropy hypothesis, which assumed the existence of stretching stress $\tau_{xx} = \mu \partial u / \partial x$ equivalent and identical to the form of shearing deformation. Consequently, the strain rate tensor can be straightforwardly included into the traditional constitutive relation, which, as a matter of fact, plays an implicitly pivotal role in establishing the N-S Eq. and has remained as a quite puzzling issue for the fluid mechanics community albeit the equations are commonly accepted and widely applied nowadays.

Towards this end, Zhu et al. (2023) first seriously questioned the isotropy issue based on the fact that the stretching stress ($\tau_{xx}$) has never been experimentally and explicitly confirmed for quite a few hundred years although the shearing stress of $\tau_{xy}$ was solidly proved by the Newton's frictional law. To effectively interrogate the issue of $\tau_{xx}$ existence via the modern DNS simulation, a corrected form of N-S Eq. (CNS) was first obtained through abandoning the isotropy hypothesis and removing the groundless stretching stress in the traditional N-S Eq. (TNS). As referenced by the fluid kinematics founded on Rortex, a pure shearing and a stretching deformation ought to be suitably identified by the off- and on-diagonal components in the strain rate tensor, respectively. Moreover, these deformation components can only be cleanly interpreted by their physical meanings and be clearly brought out in the VGT eigen-direction for the situation of either the one-real (1R) or the three-real (3R) eigenvalues. With these physical logical reasoning, the TNS was then adjusted by an added stress tensor, resulting in the CNS with the stretching stress effects due to the stretching deformations being cleanly removed in the dynamics process governed by TNS.

Given the difficulty of experimentally proving the existence of $\tau_{xx}$, the current research makes the first effort towards resolving the issue through the high-fidelity DNS applied to the fully developed turbulence in channel, i.e., a well-known benchmark for the turbulence research community. The investigation is conducted through the systematic comparisons of DNS data governed by the TNS and CNS equations to demonstrate the numerical validity and physical necessity of eliminating stretching stress effects. The comprehensive analyses include: (1) a quantitative assessment of the mean flow field, validated by the analytical wall laws with the sublayer boundaries explicitly being demarcated via the indicator function (IDF) in the entire turbulent boundary layer (TBL); (2) an in-depth evaluation of turbulence statistics, covering the Reynolds stresses, the pressure fluctuations, the components of explicitly formulated added stress tensor, the turbulent kinetic energy (TKE) budgets, and the one-dimensional premultiplied energy spectra; (3) a rigorous topological characterization of instantaneous coherent structures, utilizing the state-of-the-art Rortex and the VGT discriminant ($\Delta = 0$) in order to cleanly bring out the rotating/shearing regions and to delineate the rotational/non-rotational interface; and (4) a spatiotemporal dynamic analysis via Spectral Proper Orthogonal Decomposition (SPOD), encompassing the reduced-order decomposition and the frequency-time analysis to quantify the structural fragmentation and temporal intermittency of coherent motions. Collectively, these studies offer novel insights into fluid kinematics and dynamics, clarifying the physical limitations in traditional governing equations and capturing the more refined turbulence physics intrinsic to the fundamental fluid deformations of pure shearing and rigid rotation.

## 2 Mathematical and Physical Model of the Corrected Navier-Stokes Equations

Historically, the fluid dynamics equations, namely the N-S Eq., were established physically based on the foundation studies from the kinematic and dynamic perspectives. First of all, the fluid kinematics study has to achieve a physics-clean description of fluid fundamental deformations contained in the VGT tensor. Secondly, the corresponding dynamic responses to these deformations have to be accurately obtained in terms of their according stress generation. Within the context, the TNS equations are predicated on the C-S decomposition of VGT and on the conventional Newtonian constitutive relation with the debatable Stokes' isotropic hypothesis.

### 2.1 On the fluid kinematics

The recent fluid kinematics based on the Rortex (Liu et al. 2018) suggests that the C-S decomposition (Cauchy 1828) for a VGT ought to be further decomposed into a more fundamental level, resulting in the physics-clean tensors of rotating, shearing, and stretching (R-S-S decomposition), to accurately describe the fluid deformations with their definitive physical meanings. Moreover, clearly identifying these fundamental deformations has to be conducted in the real eigenvector directions as demonstrated in the derivation process of Rortex quantities to accurately represent vortex motions. Therefore, the fluid physics-oriented kinematics are established through further breaking up the strain rate (***A***) and vorticity (***B***) tensors in the C-S decomposition into the physics-purity level, which yields the R-S-S based decomposition with each tensor possessing its specific and clear physical meaning as follows:

$$\nabla \boldsymbol{v} = \underbrace{\boldsymbol{A}}_{Strain\ Rate} + \underbrace{\boldsymbol{B}}_{Vorticity} = \underbrace{\underbrace{\boldsymbol{SH}_{sym}}_{Symmetric\ Shearing} + \underbrace{\boldsymbol{ST}}_{Streching}}_{\boldsymbol{A}} + \underbrace{\underbrace{\boldsymbol{RO}}_{Rotation} + \underbrace{\boldsymbol{SH}_{anti-sym}}_{Anti-symmetric\ Shearing}}_{\boldsymbol{B}}$$

$$= \underbrace{\begin{bmatrix} 0 & \frac{1}{2}\varepsilon & \frac{1}{2}\varsigma \\ \frac{1}{2}\varepsilon & 0 & \frac{1}{2}\eta \\ \frac{1}{2}\varsigma & \frac{1}{2}\eta & 0 \end{bmatrix}}_{\boldsymbol{SH}_{sym}} + \underbrace{\begin{bmatrix} \lambda_1 \text{ or } \lambda_r & 0 & 0 \\ 0 & \lambda_2 \text{ or } \lambda_{cr} & 0 \\ 0 & 0 & \lambda_3 \text{ or } \lambda_{cr} \end{bmatrix}}_{\boldsymbol{ST}} + \underbrace{\begin{bmatrix} 0 & 0 \text{ or } \left(-\frac{1}{2}R\right) & 0 \\ 0 \text{ or } \left(\frac{1}{2}R\right) & 0 & 0 \\ 0 & 0 & 0 \end{bmatrix}}_{\boldsymbol{RO}} + \underbrace{\begin{bmatrix} 0 & -\frac{1}{2}\varepsilon & -\frac{1}{2}\varsigma \\ \frac{1}{2}\varepsilon & 0 & -\frac{1}{2}\eta \\ \frac{1}{2}\varsigma & \frac{1}{2}\eta & 0 \end{bmatrix}}_{\boldsymbol{SH}_{anti-sym}} \quad (2.1)$$

where ***A*** and ***B*** are the strain rate and vorticity tensors in the C-S decomposition with ***A*** being further broken up into $\boldsymbol{SH}_{sym}$ (symmetric shearing), ***ST*** (stretching) and ***B*** being decomposed into ***RO*** (rigid-rotating), $\boldsymbol{SH}_{anti\text{-}sym}$ (anti-symmetric shearing) under the corresponding real-eigenvector direction. The R-S-S decomposition can then be deemed as reaching to the fundamental level with the distinctive physics purity.

### 2.2 On the fluid dynamics

With the fluid kinematics being well-established and their physical meanings being thoroughly understood, the research in Zhu et al. (2023) then carefully revisited the classical Newtonian constitutive relation, concluding that (1) the well-known Newton frictional law only provided the evidence for a symmetric shearing deformation to generate the viscous shear stress; (2) the normal stresses induced by the stretching deformations were solely yielded by an isotropic assumption in the traditional constitutive relation that had to be removed unless being solidly proved; (3) the stress tensor ought to be symmetric due to the requirement of angular momentum conservation and therefore, only the symmetric portions in the R-S-S decomposition, namely the symmetric shearing and stretching deformations, needed to be taken into account in the new constitutive relation. Based on these fundamental knowledges and understandings, the Newtonian fluid constitutive relation can then be rewritten as $\boldsymbol{F}_{sym}=2\mu\boldsymbol{SH}_{sym}-p\boldsymbol{I}$ with the $\boldsymbol{F}_{sym}$, $\boldsymbol{SH}_{sym}$ and $\boldsymbol{I}$ being the symmetric stress, the symmetric shearing deformation and the unit tensors, where the $\mu$ and $p$ are the fluid dynamic viscosity and static pressure, respectively. The TNS are thus corrected by the $\boldsymbol{F}^{add}$ stress tensor, which cleanly cancelled out the stresses incurred by the stretching deformations in the strain rate tensor due to the Stokes' isotropic assumption. The CNS equations are formulated as follows:

$$\underbrace{\underbrace{\frac{\partial(\rho \boldsymbol{v})}{\partial t}+\nabla\cdot(\rho\boldsymbol{v}\boldsymbol{v})=\rho\boldsymbol{f}-\nabla p+\nabla\cdot\left[\mu\left(\nabla\boldsymbol{v}+\nabla\boldsymbol{v}^{\mathrm{T}}\right)\right]}_{\text{Traditional Navier-Stokes Equations (TNS)}}+\nabla\cdot\boldsymbol{F}^{add}}_{\text{Corrected Navier-Stokes Equations (CNS)}}$$

when VGT having one-real (1R) eigenvalue or flow rotating:

$$\boldsymbol{F}^{add}=-2\mu\lambda_{cr}\left\{\boldsymbol{I}-3\begin{bmatrix} r_x r_x & r_x r_y & r_x r_z \\ r_y r_x & r_y r_y & r_y r_z \\ r_z r_x & r_z r_y & r_z r_z \end{bmatrix}\right\}$$

when VGT having three-real (3R) eigenvalue*s* or flow shearing:

$$\boldsymbol{F}^{add}=-2\mu\begin{bmatrix} \sum_{i=1}^{3}\lambda_i s_{ix}s_{ix} & \sum_{i=1}^{3}\lambda_i s_{ix}s_{iy} & \sum_{i=1}^{3}\lambda_i s_{ix}s_{iz} \\ \sum_{i=1}^{3}\lambda_i s_{iy}s_{ix} & \sum_{i=1}^{3}\lambda_i s_{iy}s_{iy} & \sum_{i=1}^{3}\lambda_i s_{iy}s_{iz} \\ \sum_{i=1}^{3}\lambda_i s_{iz}s_{ix} & \sum_{i=1}^{3}\lambda_i s_{iz}s_{iy} & \sum_{i=1}^{3}\lambda_i s_{iz}s_{iz} \end{bmatrix} \tag{2.2}$$

where $\boldsymbol{v}$ is the velocity vector, $\boldsymbol{f}$ is the volume force and $p$ is the pressure. It needs to be noted that the corrected stresses of $\boldsymbol{F}^{add}$ in the CNS are determined and formulated for the two eigenvalue cases of the VGT: namely (1) the 1R-case (the rotating flow) with $\lambda_{cr}$ being the real part of the complex eigenvalue and $\boldsymbol{r} = r_x\boldsymbol{i} + r_y\boldsymbol{j} + r_z\boldsymbol{k}$ being the real eigenvector; (2) the 3R-case (the shearing flow) with the ($\lambda_i$, $i$ = 1, 2, 3) being the three eigenvalues and ($\boldsymbol{s}_i = s_{ix}\boldsymbol{i} + s_{iy}\boldsymbol{j} + s_{iz}\boldsymbol{k}$, $i$ = 1, 2, 3) being the three eigenvectors.

### 2.3 Comments on the discriminant of VGT tensor

From the fluid kinematics perspective, the VI campaign of Rortex (Liu et al. 2018) highlights the fundamental VGT eigenvalue problem, given by $\left|\nabla\boldsymbol{v}-\lambda\boldsymbol{I}\right|=0$, i.e. a cubic equation in the general form of $\lambda^3+a\lambda^2+b\lambda+c=0$. Its discriminant $\Delta$ ( $\Delta=\left[\left(2a^2-9ab+27c\right)/54\right]^2-\left[\left(a^2-3b\right)/9\right]^3$ ) then uniquely determines the rotating and shearing states of flow field and the resulting R-R-S decomposition of Eq. (2.1) yields the four basic deformation tensors with their distinct physical interpretations. Within the context, $\Delta$ is identified as the sole criterion to cleanly define the rotational/non-rotational interface (RNRI), which is quite in line with the hot-topic study of turbulent/non-turbulent interface (TNTI) in the contemporary turbulence literature (da Silva et al. 2014; Abreu et al. 2022; Zhang & Wang 2025). As a consequence, the three distinctive statuses of flow field can thus be categorized by $\Delta$ as:

(1) 1R-case ($\Delta > 0$, Rotational Region): The eigen-equation yields one real root $\lambda_r$ and a pair of complex conjugate roots $\lambda_{cr} \pm i\lambda_{ci}$. The 3D space characterized by $\Delta > 0$ is strictly identified as the rotational domain. The real eigenvector $\boldsymbol{e}_r$ defines the local rotation-axis and the local rigid-rotation strength can be explicitly calculated by $R = \boldsymbol{\omega} \cdot \boldsymbol{r} - \sqrt{(\boldsymbol{\omega} \cdot \boldsymbol{r})^2 - 4\lambda_{ci}^2}$ in Wang et al. (2019).

(2) 3R-case ($\Delta < 0$, Non-rotational Region): The equation possesses three distinct real eigenvalues ($\lambda_1$, $\lambda_2$, $\lambda_3$) corresponding to three linearly independent eigenvectors. Due to the absence of complex conjugate eigenvalues, no closed orbits or spiral topological structures can form; thus, the corresponding 3D space is the non-rotational region.

(3) 2R-case ($\Delta = 0$, Point sets with repeated roots of eigenvalues): The situation mathematically dictates the existence of algebraic repeated roots, which predominantly serves as the transitional interface partitioning the rotational and non-rotational regions or RNRI. Based on the algebraic multiplicity (AM) and geometric multiplicity (GM) of the repeated roots, the case is further sub-categorized as follows:

- AM = 2 with GM = 1 (Defective Matrix, RNRI): The subset represents the most prevalent state on the 2D iso-surface defined by $\Delta = 0$, mathematically referred to as the manifold corresponding to a 2D spatial topological geometry satisfying $\nabla\Delta(\boldsymbol{x}) \neq 0$ ($\boldsymbol{x}$ is the space coordinate vector), physically interpreted as the partition boundary between the rotation and non-rotation 3D domains or RNRI. As the matrix becomes defective (GM < AM), the eigenspace associated with the repeated roots collapses into an aligned vector $\boldsymbol{e}_{2,3}$, which physically signifies that the original rotational plane from the 1R-case smoothly folds and degenerates into a single real vector as the rotation rate $\lambda_{ci}$ decays to zero.

- AM = 2 with GM = 2 (Complete Matrix, Axisymmetric strain state): The subset corresponds to the very rare flow scenario with the two independent eigenvectors of the repeated eigenvalue, spanning a 2D eigenspace. The local topology of RNRI degenerates into a shear-free star node as put by Chong et al. (1990), which corresponds to a perfect axisymmetric strain state (Lund & Rogers 1994) in kinematics and satisfies $\nabla\Delta(\boldsymbol{x}) = 0$ with the point being named as a Singularity of Algebraic Varieties.
- AM = 3 ($\lambda_1 = \lambda_2 = \lambda_3 = 0$, Singularities): The subset stands for an extremely singular state when the VGT becomes nilpotent, corresponding to the collapse of the local velocity-gradient invariants at the degenerate boundary of the topological classification (Chong et al. 1990; Cantwell 1992). For incompressible flows ($\nabla \cdot v = 0$), if one root is zero on the $\Delta = 0$, the remaining roots must also vanish. Depending on the GM, the flow structure in the case degrades into highly singular topologies, ranging from uniform translation (GM = 3) to simple shearing flow (GM = 2), or highly degenerated shearing-only stagnation points (GM = 1).

While the mathematical condition of $\Delta = 0$ theoretically encompasses the axisymmetric strain state of AM = 2 with GM = 2 and the extreme degeneracy of AM = 3, these specific states require a delicate and exact cancellation of highly coupled spatial gradients. In a fully developed and chaotic turbulence, such as the channel turbulence in the current DNS data, the inherent strong non-normality of VGT, driven by the intensive shearing-rotating interactions, mathematically dictates that the $\Delta = 0$ state almost exclusively manifests as a defective matrix state (AM = 2 with GM = 1). Consequently, in the practical post-processing of DNS data, the explicit condition of $\Delta = 0$ is topologically and physically equivalent to the exact RNRI dictated by the AM = 2 with GM = 1 state. Based on the conclusion, the geometric presentations of RNRI are further visualized and analyzed in **Section 4.2** along with the Rortex iso-surfaces at certain threshold and the vortex core line, which provides the most comprehensive and unique topological description of vortical structures.

### 2.4 Comments on the stretching deformation and correction stress tensor in CNS

In continuum mechanics, the local stress state of a fluid element is fully characterized by the Cauchy stress tensor, $\tau_{ij}$, where the index $i$ denotes the surface normal vector and $j$ indicates the direction of the force component. Although the stretching stress represented by $\tau_{xx} = \mu\,\partial u/\partial x$ under the physical coordinates is quite difficult to be observed or detected directly in experiments, its physical meaning can evidently be interpreted as the stress generated by the deformation in the direction of fluid motion, which is found quite similar to the scenario of Hooke's law ($F = -kx$) for a spring, particularly in the sense that the traction $\tau_{xx}$, the velocity component $u$, and the surface-normal direction $x$ in $\tau_{xx} = \mu\,\partial u/\partial x$ are strictly collinear as depicted in Fig. 1(a). On the contrary, for the shearing case of $\tau_{yx} = \mu\,\partial u/\partial y$, the traction $\tau_{yx}$ and velocity component $u$ are perpendicular with the surface-normal direction $y$ as exhibited in Fig. 1(b). Therefore, it is quite reasonable to argue that $\tau_{xx}$ can be regarded as analogous to an elastic-like stress model applied to fluid instead of solid. Within the context, the traditional Newtonian fluid model is, as a matter of fact, a viscous fluid mixed with an elastic-stress model when subjected to stretching deformation, which is somewhat contradictory to the common knowledge on fluid, at least for water or air, and consequently leads to the controversial issue. The observation and inference therefore provide a legitimate reason to revamp the conventional constitutive relation for Newtonian fluid and to remove the according stretching stresses from TNS to persist the fluid physical soundness in the dynamic model.

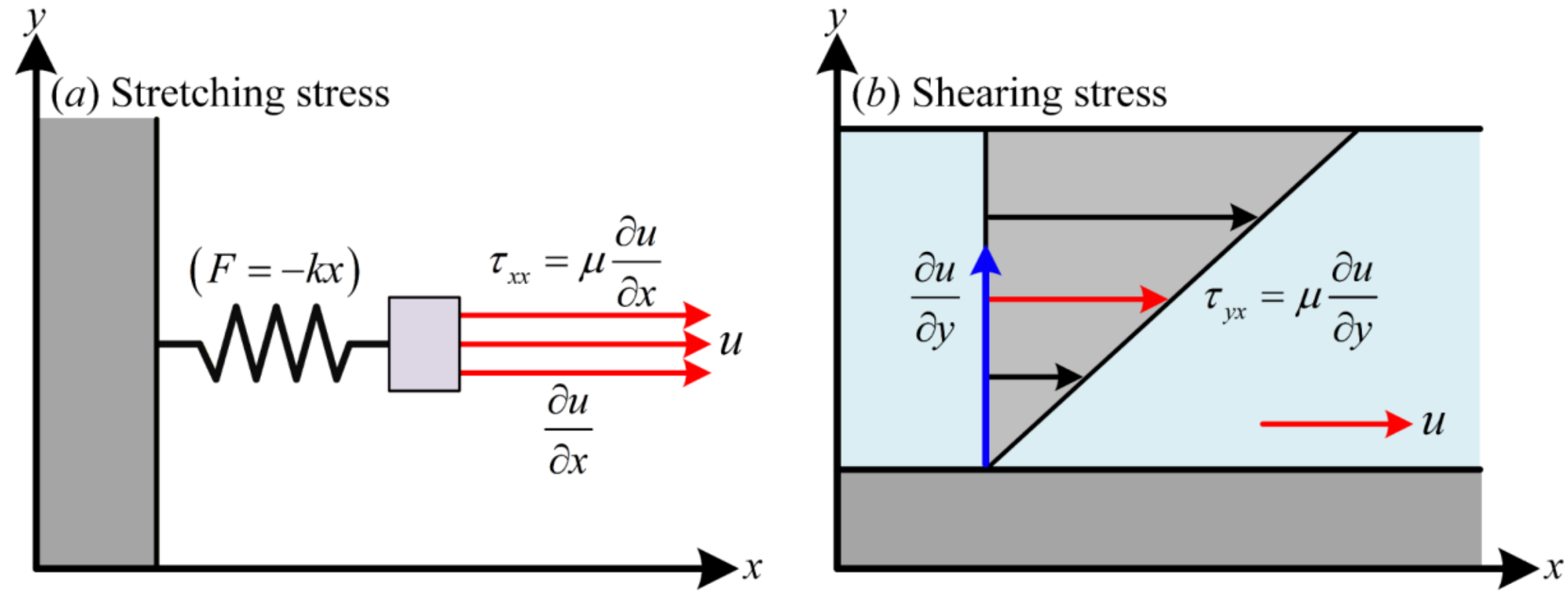

FIG. 1. Schematic of (a) stretching and (b) shearing stress configurations in fluid (Newton friction law) and solid (Hooke elastic law) scenarios.

Moreover, the governing equations, i.e., Eq. (2.2), are derived through the R-R-S decomposition (kinematics renovation) and removal of the stretching stresses in the conventional constitutive relation (dynamics correction). The inclusion of $\boldsymbol{F}^{add}$ in Eq. (2.2) explicitly suggests that the correction to TNS has to be nonlinear and dependent on the eigenvalue status of VGT, which is incurred by the necessity to cleanly identify the stretching stress (***ST***) by the diagonal component in the tensor under the eigenvector direction and the rotation transform between the eigendirection and the physical coordinate direction. Unlike the turbulence models in the current RANS and LES enabled by the eddy viscosity concept based on the analogy between fluid and molecular motions, the $\boldsymbol{F}^{add}$-tensor correction is rooted in the fluid kinematics and dynamics with the strict mathematical foundation and rigorous physical logic demonstrated in **Sections 2.1** and **2.2**. Therefore, it is expected that directly solving the corrected equations of CNS and comparing the solution with the existing DNS data based on TNS is expected to capture more precise fluid physics and to bring out more accurate vortex structures in turbulence through revamping the fluid physical soundness in the dynamic model. The existing DNS or LES data based on TNS may, high-likely to some extent, twist the correct fluid physics because of either the groundless eddy viscosity assumption or the imaginary stretching stresses due to the Stokes' isotropic hypothesis.

It is worth noting that the form of CNS possesses two types of nonlinearities in fluid dynamics, namely (1) the nonlinearity caused by the convection terms in the product form algorithm of velocity components, with one of its well-known effects of the Reynolds stresses in TNS; (2) the nonlinearity induced by the removal of stretching stresses in CNS, which is introduced by the $\boldsymbol{F}^{add}$ terms and is proportional to the inverse of Re when the CNS is nondimensionalized. Within the context, the nonlinearity incurred by $\boldsymbol{F}^{add}$ is referred to as the viscous nonlinearity since these terms are essentially related to the viscosity in CNS. Distinguishing the two types of nonlinearities is important in the following DNS data analyses based on both TNS and CNS.

## 3 Direct Numerical Simulations of Turbulent Channel Flow based on TNS and CNS

### 3.1 Computational model, boundary condition, domain and grid setup

The DNS were performed to resolve the incompressible fully developed turbulent channel flow based on the governing equations of both TNS and CNS, which were coupled to the continuity equations in the non-dimensional form, as presented in Eqs. (3.1), (3.2), and (3.3), respectively,

$$\nabla \boldsymbol{v}^{+} = 0 \tag{3.1}$$

$$\frac{\partial \boldsymbol{v}^{+}}{\partial t} + (\boldsymbol{v}^{+} \cdot \nabla)\boldsymbol{v}^{+} = -\nabla p^{+} + \frac{\overline{\partial P}}{\partial x} + \frac{1}{Re_{\tau}}\nabla^{2}\boldsymbol{v}^{+} \tag{3.2}$$

$$\frac{\partial \boldsymbol{v}^+}{\partial t} + (\boldsymbol{v}^+ \cdot \nabla)\boldsymbol{v}^+ = -\nabla p^+ + \frac{\overline{\partial P}}{\partial x} + \frac{1}{Re_\tau}\nabla^2 \boldsymbol{v}^+ + \nabla \cdot \boldsymbol{F}^{add+} \tag{3.3}$$

where $\boldsymbol{v}^+$ denotes the velocity vector, with the components of ($u^+$, $v^+$, $w^+$) corresponding to the streamwise($x$), wall-normal ($y$), and spanwise ($z$) directions; $t$ represents time, $\overline{\partial P/\partial x}$ is the mean streamwise pressure gradient driving the channel turbulence and $\nabla p^+$ is the incremental pressure gradient due to turbulence with the vector operator $\nabla$ being defined as $\nabla = \partial/\partial x\, \boldsymbol{i} + \partial/\partial y\, \boldsymbol{j} + \partial/\partial z\, \boldsymbol{k}$; the added stress tensor $\boldsymbol{F}^{add+}$ is defined for the 1R and 3R-situations in Eq. (2.2); the ensemble-averaged frictional velocity ($\bar{u}_\tau$) and half of the channel width ($\delta$) are applied as the characteristic velocity and length, respectively; the $\bar{u}_\tau$-based Reynolds number $Re_\tau$ is thus calculated by $Re_\tau = \bar{u}_\tau \delta/\nu$ with $\nu$ being the fluid kinematic viscosity.

In the following discussions, the mean quantities are represented by the capital letters and the fluctuations by upper apostrophe (e.g., $u = U + u'$, where $U = \langle u \rangle$ with $\langle \bullet \rangle$ denotes the ensemble average defined as the statistical averaging over both time and space of the periodic direction). The quantities normalized by the wall frictional units are denoted by the superscript '+', based on the kinematic viscosity $\nu$ and the friction velocity of $\bar{u}_\tau$ defined as $\bar{u}_\tau = \sqrt{\bar{\tau}_w/\rho}$, where $\bar{\tau}_w$ is the ensemble-averaged wall shear stress and $\rho$ is the fluid density.

The periodic boundary conditions were prescribed in the streamwise ($x$) and spanwise ($z$) directions with the Fourier series expansions being applied. In the wall-normal direction ($y$), the no-slip velocity conditions were imposed on the wall surface and the non-uniform grid with a near-wall refinement was applied to ensure an adequate resolution in the viscous sublayer. A constant mass flux throughout the channel was maintained by the streamwise mean pressure gradient ($\overline{\partial P/\partial x}$), which was equal to a constant under the chosen characteristic velocity of $\bar{u}_\tau$ and the length of $\delta$.

The computational domain was set as the rectangular cuboid in Fig. 2 with the lengths of $L_x = 2\pi\delta$ in streamwise, $L_y = 2\delta$ in wall-normal and $L_z = \pi\delta$ in spanwise. The flow field was discretized based on the uniform grids of $N_x$ = 515 and $N_z$ = 515 in the periodic directions due to the requirement of the Fourier transformation and the non-uniform grids of $N_y$ = 515 were applied in the wall-normal direction.

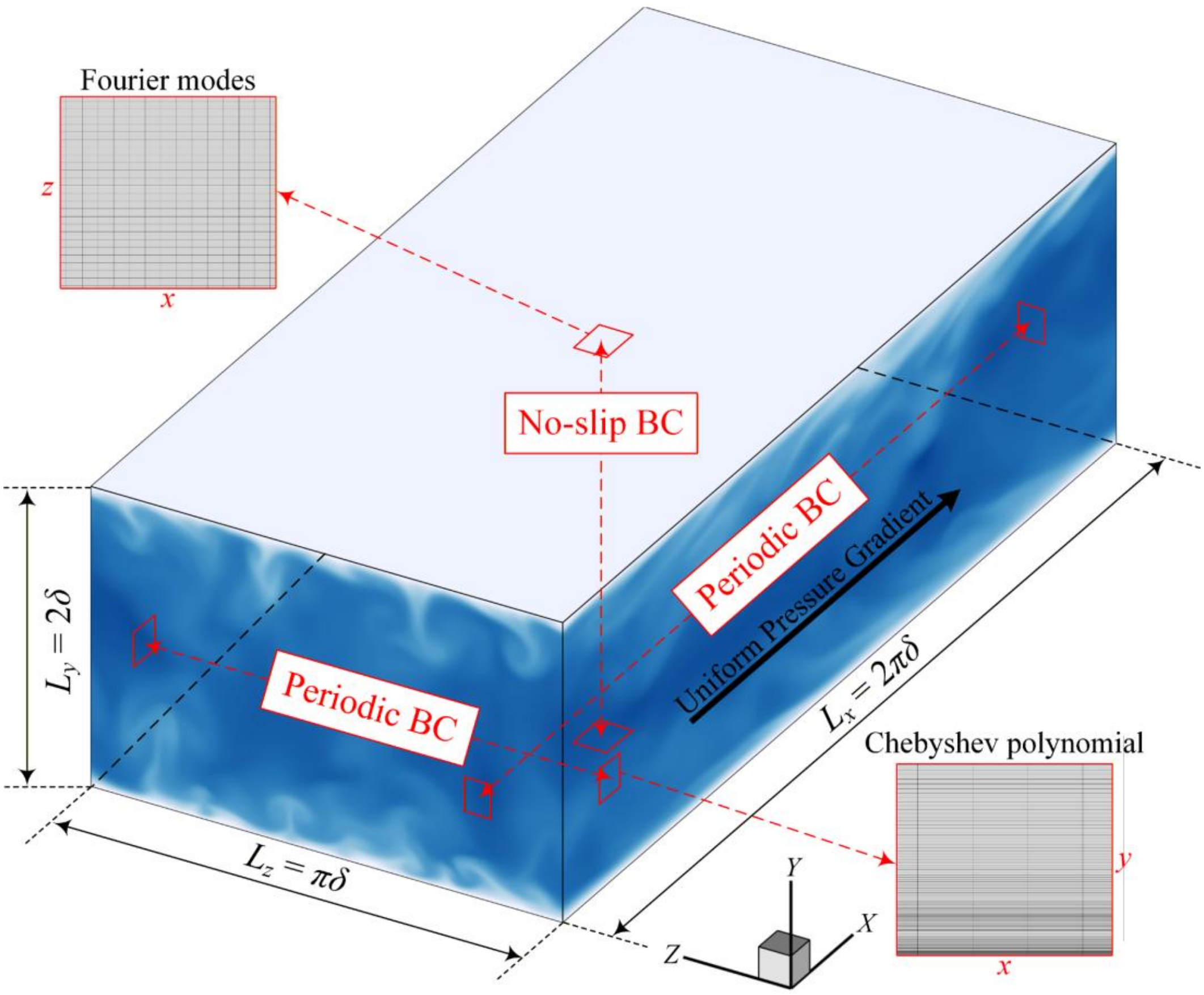

FIG. 2. Schematic of the computational domain for the DNS of turbulent channel flow.

The DNS simulation was performed at $Re_\tau = 550$, matching with the $Re_\tau$ reported in Lee & Moser (2015), which was selected as the benchmark for the accuracy estimation and algorithm validation, since these data were demonstrated to surpass the established standards for wall-bounded turbulence DNS through achieving the statistical uncertainties of less than 1%. The $\bar{u}_\tau$-based grid intervals of $\Delta(x, y, z)^+$ are provided in Table I for measuring and comparing the spatial resolutions.

TABLE I. Summary of the simulation parameters for the DNS based on the CNS and TNS. The LM dataset refers to the DNS data of Lee & Moser (2015), serving as the benchmark for accuracy estimation and algorithm validation. Here, $\Delta y_w$ and $\Delta y_c$ denote the grid spacings at the wall and centerline, respectively, and $Tu_\tau/\delta$ represents the total large-eddy turnover time (LETOT) for collecting the turbulence statistics.

| Name | $Re_\tau$ | $L_x/\delta$ | $L_z/\delta$ | $L_y/\delta$ | $N_x$ | $N_z$ | $N_y$ | $\Delta x^+$ | $\Delta z^+$ | $\Delta y_w^+$ | $\Delta y_c^+$ | $Tu_\tau/\delta$ |
|---|---|---|---|---|---|---|---|---|---|---|---|---|
| CNS | 550 | $2\pi$ | $\pi$ | 2 | 515 | 515 | 515 | 6.8 | 3.4 | 0.023 | 6.8 | 30 |
| TNS | 550 | $2\pi$ | $\pi$ | 2 | 515 | 515 | 515 | 6.8 | 3.4 | 0.023 | 6.8 | 30 |
| LM | 550 | $8\pi$ | $3\pi$ | 2 | 1536 | 1024 | 384 | 8.9 | 5.0 | 0.019 | 4.5 | 13.6 |

For the present grid, the streamwise ($x$) and spanwise ($z$) resolutions are finer than those of the LM benchmark. Although the wall-normal ($y$) spacing is slightly coarser, it remains within the same order of magnitude. The configuration is deemed as tolerable since the LM data is notably over-resolved with a grid sufficient to satisfy the accuracy requirements up to $Re_\tau = 1000$. Furthermore, the current wall-normal resolution is better than the one in the classic benchmark at $Re_\tau = 590$ by Moser et al. (1999). Consequently, the grid adopted in the present study is considered adequate to ensure the accuracy required for a DNS simulation. On the other hand, given the limitation of computing power in the study, the domain size in the current DNS investigation has to be significantly reduced to match the grid resolution with the existing DNS benchmark, namely four times lower in streamwise and three times lower in spanwise than those in the LM data, which may be one of the key factors causing the discrepancies (majorly under-resolving the longer-wave length energies) in the data comparisons presented in the following sections.

### 3.2 Discretization schemes, algorithms and solution methods

Time integration was performed using a semi-implicit approach with the second-order explicit Adams–Bashforth scheme being applied to the nonlinear convective and added stress terms, and the second-order implicit Adams–Moulton scheme being adopted for the diffusive terms in both TNS and CNS. The spatial discretization employed the second-order finite volume method on a staggered grid. To satisfy stability and accuracy requirements, the Minmod Total Variation Diminishing (TVD) scheme was utilized for the reconstruction of interfacial fluxes. The continuity and momentum equations were decoupled via the fractional step method of Kim & Moin (1985). To enhance computational efficiency, the solution method based on Fast Fourier Transform (FFT) was applied in the streamwise ($x$) and spanwise ($z$) directions by taking advantage of periodic boundary conditions. After attaining the fully developed and statistically stationary state, the DNS simulations were conducted at a fixed time step of $\Delta t = 1.0 \times 10^{-4}$, ensuring the Courant–Friedrichs–Lewy (CFL) number remained below 0.5 for a stable time advancement, and the turbulence statistics were thereafter accumulated over a further 30 large eddy turnover times.

### 3.3 Turbulence statistics stationarity verification

The total shear stress profile is examined to verify the statistical stationarity of the simulated turbulence. In a fully developed, statistically stationary turbulent channel flow governed by the TNS equations, the total stress, comprising the sum of the Reynolds shear stress and the mean viscous stress, must contribute to a linear distribution (Pope 2000) across the channel to satisfy the momentum conservation as presented by Eq. (3.4). However, regarding the CNS, the corresponding shear-correction term in $\boldsymbol{F}^{add}$ also plays an important role in the total stress balance to maintain the strict linearity, as formulated in Eq. (3.5). Consequently, any deviation from the theoretical linear profile serves as a direct metric for quantifying the momentum equation residual, as illustrated in Fig. 3.

$$\frac{\partial U^+}{\partial y^+} - \langle u'v' \rangle^+ = 1 - \frac{y}{\delta} \tag{3.4}$$

$$\frac{\partial U^+}{\partial y^+} - \langle u'v' \rangle^+ - \left\langle F_{xy}^{add} \right\rangle^+ = 1 - \frac{y}{\delta} \tag{3.5}$$

As illustrated in Fig. 3, the overall discrepancy between the analytical linear profile and the computed total stress profile remains on the same order of magnitudes across all the simulations. The maximum simulation residuals are both within the range of 0.002 for both CNS and TNS, even better than the maximum residual in LM data. The residual peaks tend to occur at the location around $y^+ \approx 30$, corresponding to the transition between the buffer and logarithmic layers where both the viscous and Reynolds shear stresses (viscous and convection nonlinearities) remain dynamically significant. Consequently, the intense local fluctuations in the total stress balance induce the elevated residuals with the localized peaks in the time-averaged statistical profiles. Conclusively, the quite high confinement level to the theoretical linearity relation confirms that all the simulations well achieve the satisfactory statistically stationary state in the channel turbulence. Furthermore, the subsequent analysis unveils a reduction of the Reynolds shear stress $-\langle u'v' \rangle^+$ in the CNS model as highlighted by Fig. 5(b), which proves to serve as a dynamic compensation for the increase of $-\left\langle F_{xy}^{add} \right\rangle^+$ observed in Fig. 7(d), and thereby ensures the streamwise momentum conservation governed by Eq. (3.5). In other words, whilst the correction incurred by $\boldsymbol{F}^{add}$ modifies the internal stress distribution, the CNS model still fundamentally preserves the total stress balance inherent in the nature of channel turbulence.

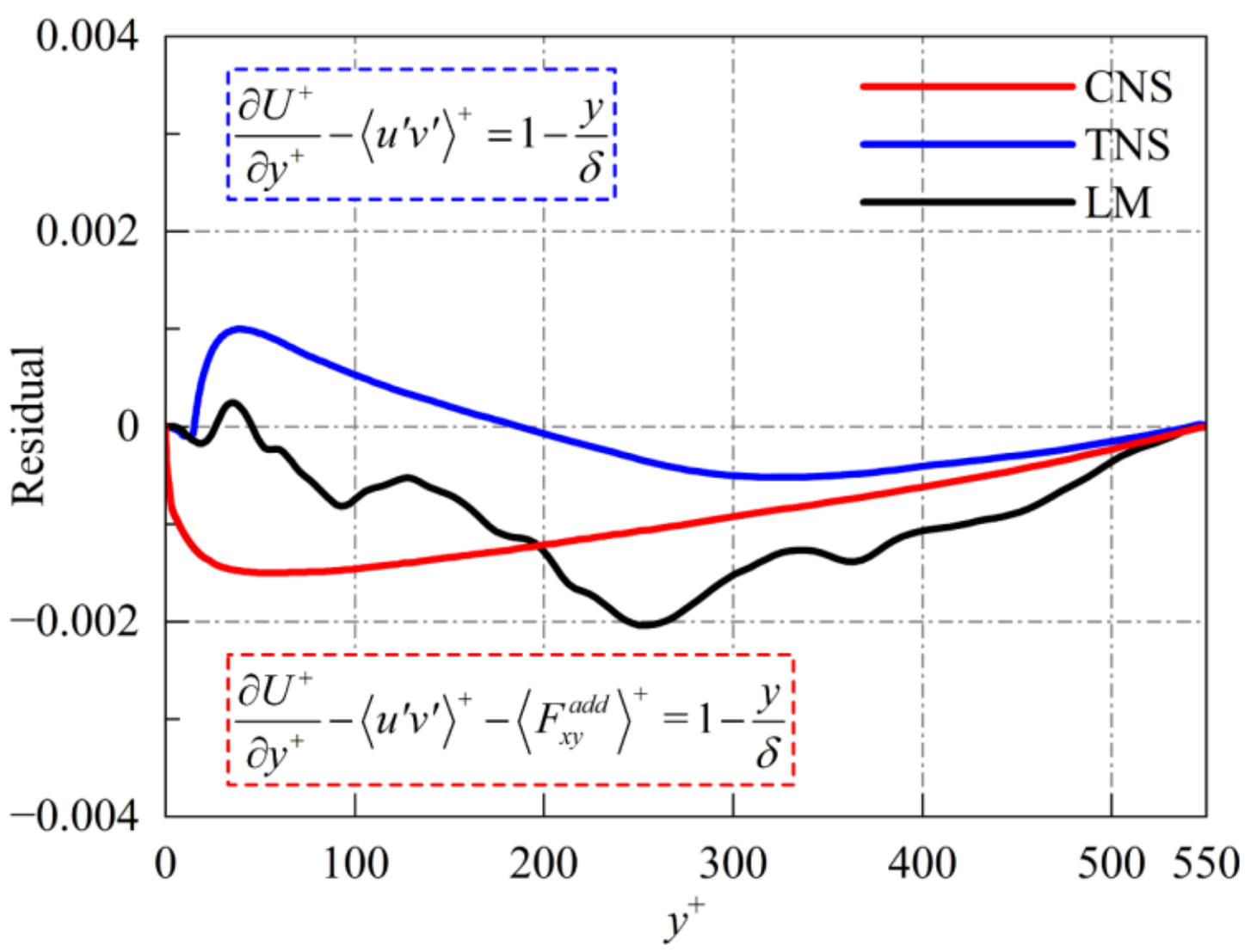

FIG. 3. Statistical stationarity proofs of the simulations with the CNS, TNS, and LM results being denoted by the red, blue, and black lines, respectively.

### 3.4 Spectral proper orthogonal decomposition method

The extraction of coherent structures from high-dimensional turbulent flow data is pivotal to uncover the underlying physical mechanisms of wall-bounded turbulence. While the classical data-driven modal decomposition techniques, such as the Proper Orthogonal Decomposition (POD; Lumley 1967; Sirovich 1987; Berkooz et al. 1993), provide an energy-optimal basis, the method frequently amalgamates structures occurring at disparate frequencies, thereby obfuscating the true spatiotemporal dynamics. Conversely, the Dynamic Mode Decomposition (DMD; Schmid 2010) isolates the single-frequency modes but lacks the inherent energy optimality and orthogonality properties of POD. To bridge the gap, the current study employs the SPOD technique. Originally conceptualized by Lumley (1970) and recently revitalized and theoretically unified with resolvent analysis (Towne et al. 2018), the SPOD yields modes that are both optimal in an energy sense and strictly monochromatic. In the context of the present study on channel turbulence DNS data, the SPOD suitably offers a distinct advantage by enabling the decomposition of broadband turbulence into the spectrally clean modes. For statistically stationary data, the modes are computed in the frequency domain by solving an eigenvalue problem for the cross-spectral density (CSD) tensor, which represents the Fourier transform of the two-point space–time correlation. To ensure a robust convergence of the spectral modes, the Welch's method (Welch 1967) was adopted as outlined by Schmidt & Colonius (2020). The data sequence of $nt = 3000$ snapshots, sampled at a time step of $\Delta t = 1 \times 10^{-2}$ yielding a sampling frequency $f_s = 100$ and a Nyquist limit of $f_{max} = 50$, is segmented into $n_{blk} = 22$ overlapping blocks. Each block comprises $n_{fft} = 256$ snapshots, providing a sufficient frequency resolution to capture the low-frequency, large-scale coherent structures characteristic of broadband channel turbulence. The SPOD brings out the modes that are optimal with respect to a spatial inner product defined by a positive-definite Hermitian weight matrix $\boldsymbol{W}$. A component-wise weighted 2-norm was employed as a numerical quadrature of the inner product within a continuous Hilbert space, explicitly incorporating the volume of individual computational cells to compensate for the non-uniform mesh distribution in the wall-normal direction. To mitigate the spectral leakage, each block was pre-multiplied by a symmetric Hamming window (Harris 1978). A 50% overlap ($n_{ovlp}$ = 128) between successive blocks was implemented to recover information otherwise likely being attenuated at the window boundaries, which treated each block as a statistically independent realization under the ergodic hypothesis. Following the discrete Fourier transform (DFT) of the windowed blocks, the data are reorganized by frequency. The SPOD modes, $\boldsymbol{\Phi}$, and their corresponding modal energies, $\lambda$, are formally defined as the eigenvectors and eigenvalues of the CSD matrix. By construction, the SPOD modes are orthogonal in the space–time inner product. All SPOD parameters are summarized in Table II.

TABLE II. Configuration of the SPOD parameters. Note that these parameters are application-dependent rather than intrinsic constants of the standard SPOD framework; the presented values are selected as the optimal configuration for the current study and all quantities are normalized.

| Parameters | Value |
|---|---|
| Total sample size $n_t$ | 3000 |
| Sampling time step $\Delta t$ | $1\times10^{-2}$ |
| DFT window $n_{fft}$ | 256 |
| Overlap $n_{ovlp}$ | 128 |
| Blocks $n_{blk}$ | 22 |
| Weight matrix $\boldsymbol{W}$ | Weighted 2-Norm |
| Window | Hamming |
| Frequency-time analysis | Convolution-based approach |

## 4 DNS Data Analyses and Discussions

### 4.1 Statistics of fully developed turbulent channel flow

#### *4.1.1 Mean flow field*

To validate the current DNS data, the reliable experimental measurements are indispensable. In this regard, den Toonder & Nieuwstadt (1997) utilized the laser doppler velocimetry (LDV) to measure a fully developed turbulent pipe flow at $Re_\tau \approx 520$, while Hussain & Reynolds (1975) employed the hot-wire anemometry (HWA) to investigate a fully developed turbulent channel flow at $Re_\tau \approx 640$. These measurement datasets serve as the trustworthy experiments for evaluating the three sets of DNS data at $Re_\tau = 550$. The line styles and symbols selected in subsequent figures are summarized in Table III.

TABLE III. Summary of the comparative datasets with the $Re_\tau$ being calculated by the channel half-width and the respective friction velocity as the characteristic quantities.

| Source | Type | Method | Model | $Re_\tau$ | Glyph |
|---|---|---|---|---|---|
| Current study | DNS | CNS | Channel | 550 | —— |
| Current study | DNS | TNS | Channel | 550 | —— |
| Lee & Moser (2015) | DNS | TNS | Channel | 550 | —— |
| den Toonder & Nieuwstadt (1997) | EXP | LDV | Pipe | 520 | 🟩 |
| Hussain & Reynolds (1975) | EXP | HWA | Channel | 640 | ● |

Figure 4(a) presents the mean streamwise velocity profiles across all the datasets. In the viscous sublayer, all the data align well with the linear law, exhibiting the mean relative deviations of less than 1%. The agreement is attributed to the dominance of viscous shearing stress in the immediate near-wall region of $y^+ \leq d_{s1}^+$, which makes the effects of viscous stretching stresses negligible. In the transition region of $d_{s1}^+ \leq y^+ \leq d_{s2}^+$, the nonlinear effect of Reynolds stress emerges albeit the viscous shear stress still dominates. Within the buffer layer ($d_{s2}^+ \leq y^+ \leq d_{s3}^+$), the viscous shearing, Reynolds stresses and $\boldsymbol{F}^{add}$ terms are of comparable magnitudes when the velocity profile transits from the nearly-linear to the semi-logarithmic profiles. The traditional law of the wall postulates a semi-logarithmic variation with the wall distance for the mean streamwise velocity in semi-log sublayer ($d_{s3}^+ \leq y^+ \leq d_{s4}^+$), generally expressed as

$$U^+ = \frac{1}{\kappa}\log y^+ + B, \tag{4.1}$$

where $\kappa$ is the von Kármán constant and $B$ is the intercept. The values of $\kappa$ are determined by the CNS,

TNS, and LM datasets at 0.458, 0.424, and 0.427, respectively, through the indicator function in Fig. 4(b).

As illustrated in Fig. 4(a), the mean streamwise velocity profile of the CNS demonstrates an excellent agreement with both the LM and experimental data, whereas the TNS exhibits a noticeable overshoot. On average, the CNS profile is 6.8% lower than that of the TNS. The discrepancy arises because the convective nonlinearity becomes a dominant factor. Consequently, the mean velocity profile is governed primarily by the nonlinear convection, which physically manifests as the Reynolds stresses, particularly the shear component $-\langle u'v'\rangle^+$, as shown in Fig. 5(b). The profile difference between the TNS results and LM data is mainly attributed to the numerical dissipation inherently contained in the low-order differencing schemes (second-order TVD scheme) based on the finite volume method in the current TNS computation, which is an aspect warranting further investigation via utilizing a high-order scheme as applied in the LM data. Therefore, the present CNS data indirectly justify the necessity of the TNS correction, demonstrating its capability to yield the higher fidelity data in the regimes dominated by the nonlinear Reynolds stresses, despite that the lower-order scheme was employed.

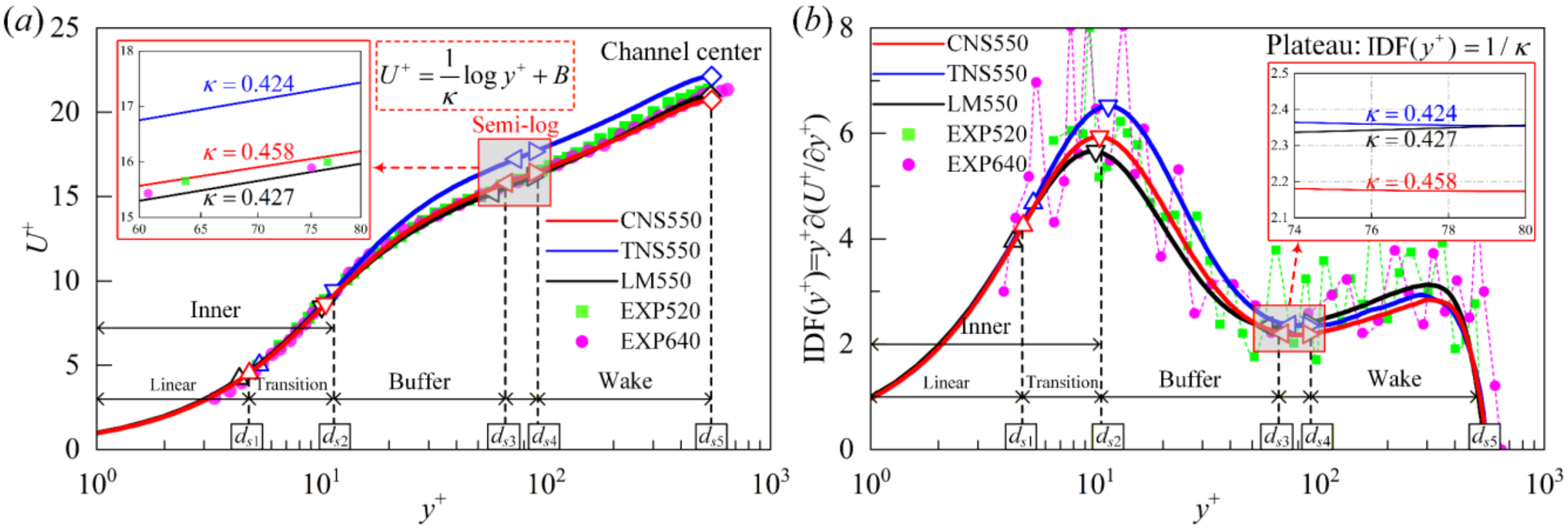

FIG. 4. Profiles of the mean streamwise velocity and log-law indicator function, with lines and symbols defined in Table III. The distinct sublayers in the law of the wall are quantitatively demarcated by the characteristic points $d_{si}^+$ ($i = 1-5$), with a magnified view of the semi-log region highlighted in the inset.

To further accurately evaluate the velocity profiles, the indicator function (IDF; Schlatter & Örlü 2010; Österlund 1999) is introduced, which takes the form as the product of the wall-normal coordinate and mean velocity gradient:

$$\mathrm{IDF}(y^+) = y^+ \frac{\partial U^+}{\partial y^+} \tag{4.2}$$

Theoretically, the IDF is expected to exhibit a constant plateau at $1/\kappa$ within the logarithmic sublayer. Figure 4(b) compares the IDF profiles across all the datasets. As highlighted in the magnified inset, the logarithmic region clearly manifests as a distinct horizontal plateau, precisely aligning with the theoretical value of $1/\kappa$.

To precisely define and quantitatively determine the boundaries of each near-wall sublayer, the sublayer boundaries were identified by the characteristic points on the IDF. Thereby, the corresponding scaling laws (Wang et al. 2022) were successfully discovered and established with their analytical formulations. The sublayers are uniformly and quantitatively defined within the intervals $d_{s(i-1)}^+ \le y^+ \le d_{si}^+$ (where $d_{s0}^+ = 0$). Here, $i = 1$ corresponds to the viscous sublayer, $i = 2$ to the inner transition layer, $i = 3$ to the buffer layer, $i = 4$ to the log-law region, and $i = 5$ to the wake zone. Furthermore, the inner layer is formed by merging the viscous sublayer and the inner transition layer. For the entire inner layer ($0 \le y^+ \le d_{s2}^+$),

$$U^+ = y^+ \left\{1 - \varepsilon_1^+ \left[1 - e^{-y^+/D_1^+}\right]\right\}, \tag{4.3}$$

where $\varepsilon_1^+$ represents the damping strength and $D_1^+$ stands for the wall-normal distance beyond which the damping becomes effective. For the buffer layer ($d_{s2}^+ \le y^+ \le d_{s3}^+$),

$$U^+ = U^+(d_{s3}^+) + \ln\left(\frac{y^+}{d_{s3}^+}\right)\left\{\frac{1}{\kappa} - \varepsilon_2^+\left[1 - e^{-\ln\left(y^+/d_{s3}^+\right)/D_2^+}\right]\right\}. \tag{4.4}$$

where $\varepsilon_2^+$ and $D_2^+$ play the same roles as $\varepsilon_1$ and $D_1^+$ in (4.3). For the log-law region ($d_{s3}^+ \leq y^+ \leq d_{s4}^+$), the classical logarithmic law (Eq. 4.1) is retained. Given the symmetry of channel turbulence at the channel centerline, the wake law is naturally treated as an extension of the log-law.

Table IV summarizes the TBL sublayer characteristics including the boundary points and control parameters for all the cases in the current study. Notably, the boundaries of each sublayer in the CNS data significantly shift toward the wall relative to those in the TNS data, which means a bigger velocity gradient and a thinner TBL near wall due to the viscous nonlinearity of $\boldsymbol{F}^{add}$, as presented in Fig. 7. Within the inner layer, the upper boundary of the viscous sublayer ($d_{s1}^+$) for the CNS moves closer to the wall by 9.6% comparing to the TNS results. Similarly, for the upper boundary of the inner layer ($d_{s2}^+$), the CNS data precede the TNS ones by 8.3%. The damping strength $\varepsilon_1^+$ and the characteristic thickness scale $D_1^+$ are also substantially lower in the CNS than those in the TNS. The removal of stretching stresses and the consequential viscous nonlinearity facilitates the excitation of Reynolds shear stress at a nearer distance from the wall, leading to an earlier ending of the viscous sublayer and thereby compacting the momentum exchange in the inner layer. Compared to the TNS data in the buffer layer, the upper boundary $d_{s3}^+$ in the CNS cases is shifted towards the wall by 10.8%, resulting in a significant reduction in the buffer layer thickness and the markedly lower control parameters ($\varepsilon_2^+$, $D_2^+$). The reduction is attributed to the removal of stretching stresses and the resulting viscous nonlinearity, which mitigates the hysteresis effect in the turbulence energy cascade and enables a faster transition from the inner to the log-law region. While the upper boundaries of the log-law region ($d_{s4}^+$) and the wake region ($d_{s5}^+$) remain almost consistent across all the models, the log-law plateau ($d_{s3}^+ \leq$ $y^+ \le d_{s4}^+$) in the CNS expands by about 154% relative to the TNS due to the wall-ward shift of $d_{s3}^+$. Regarding the control parameters in the log-law region, the von Kármán constant $\kappa$ increases from 0.424 in the TNS to 0.458 in the CNS, while the intercept $B$ decreases from 6.965 to 6.248. Note that the values are strictly obtained from the IDF($y^+$) of the plateau region and the plateau region is relatively short at $Re_\tau = 550$. Therefore, the fitted $\kappa$ may be subject to numerical uncertainty. The experimental data of den Toonder & Nieuwstadt (1997) presents a good fit with ($\kappa = 0.4$, $B = 5.5$), whereas the experiment by Hussain & Reynolds (1975) shows a better agreement with Coles' log law ($\kappa = 0.41$, $B = 5.0$; Coles 1956). However, it should be noted that, limited by the experimental measurement accuracy, neither was considered as the exact parameters since these values were merely obtained through the data-fitting comparisons. The $B$ value in CNS is closer to the experimental data, but the $\kappa$ value is slightly higher, which might be attributed to the lower-*Re* factor. The elevation in $\kappa$ signifies an increased slope of the mixing length in the eddy-viscosity assumption, suggesting that the momentum exchange is driven entirely by the pure shearing, given that the stretching stress is totally removed, and the rigid rotation achieves a higher macroscopic mixing efficiency compared to the TNS.

TABLE IV. Locations of the TBL sublayer boundary points $d_{si}^+$ ($i = 1 - 5$) and the associated control parameters of ($\varepsilon_1^+$, $D_1^+$, $\varepsilon_2^+$, $D_2^+$, $\kappa$, $B$).

| Case | $d_{s1}^+$ | $d_{s2}^+$ | $d_{s3}^+$ | $d_{s4}^+$ | $d_{s5}^+$ | $(\varepsilon_1^+, D_1^+)$ | $(\varepsilon_2^+, D_2^+)$ | $(\kappa, B)$ |
|---|---|---|---|---|---|---|---|---|
| CNS | 4.78 | 10.45 | 68.14 | 81.68 | 544.50 | (0.114, 8.457) | (1.945, 3.169) | (0.458, 6.248) |
| TNS | 5.29 | 11.39 | 76.36 | 81.68 | 544.50 | (0.119, 9.765) | (2.262, 3.274) | (0.424, 6.965) |
| LM | 4.32 | 10.03 | 58.00 | 81.68 | 544.50 | (0.056, 7.520) | (4.006, 5.486) | (0.427, 5.711) |

*4.1.2 Turbulent Reynolds stress*

Figure 5 presents the profiles of the Reynolds stress components obtained from the DNS simulations of fully developed turbulent channel flow. The streamwise Reynolds normal stress profile, $\langle u'^2 \rangle^+$, exhibits a characteristic inner peak. As given in Fig. 5(a), the CNS predictions demonstrate an excellent agreement with both the LM database and experimental measurements, yielding a 13% reduction in the peak magnitude relative to the TNS baseline. The attenuation stems from the exclusion of the stretching-induced viscous stresses due to the corrections in CNS, which consequently mitigates the deformation damping exerted on the fluid elements, as evidenced by the increased dissipation term in Fig. 8(b). The reduction promotes the breakdown of vortical structures into more fragmented zones (Figs. 10 and 11), thereby accelerating the energy redistribution from the streamwise-dominant streaks to the wall-normal and spanwise fluctuations, as shown in Figs. 5(c) and (d) with the elevated $\langle v'^2 \rangle^+$ and $\langle w'^2 \rangle^+$. Furthermore, the location of the CNS peak shifts closer to the wall by 8%. Given that the peak location typically indicates the core of buffer layer, such a wall-ward shift implies a thinning viscous sublayer and a spatial compression of near-wall scales in the CNS prediction.

Figure 5(b) compares the profiles of Reynolds shear stress, $-\langle u'v' \rangle^+$, from all the datasets. These profiles are quite close to each other, and the peak magnitude of CNS is only 2% lower than that of TNS, which suggests that the exclusion of the stretching-induced viscous stresses leads to a marginal decrease in the dynamic coupling between the streamwise and wall-normal velocity fluctuations. The decrease can also be explained by the dynamic compensation made to maintain the conservation of momentum in Eq. 3.5, due to the arising and growing of $-\left\langle F_{xy}^{add} \right\rangle^+$ in Fig. 7. Furthermore, the peak location consistently shifts wall-ward by 9%. Given that the shear stress peak aligns with the region of maximum turbulence production, the spatial shift indicates that the active zone of turbulence generation moves closer to the wall.

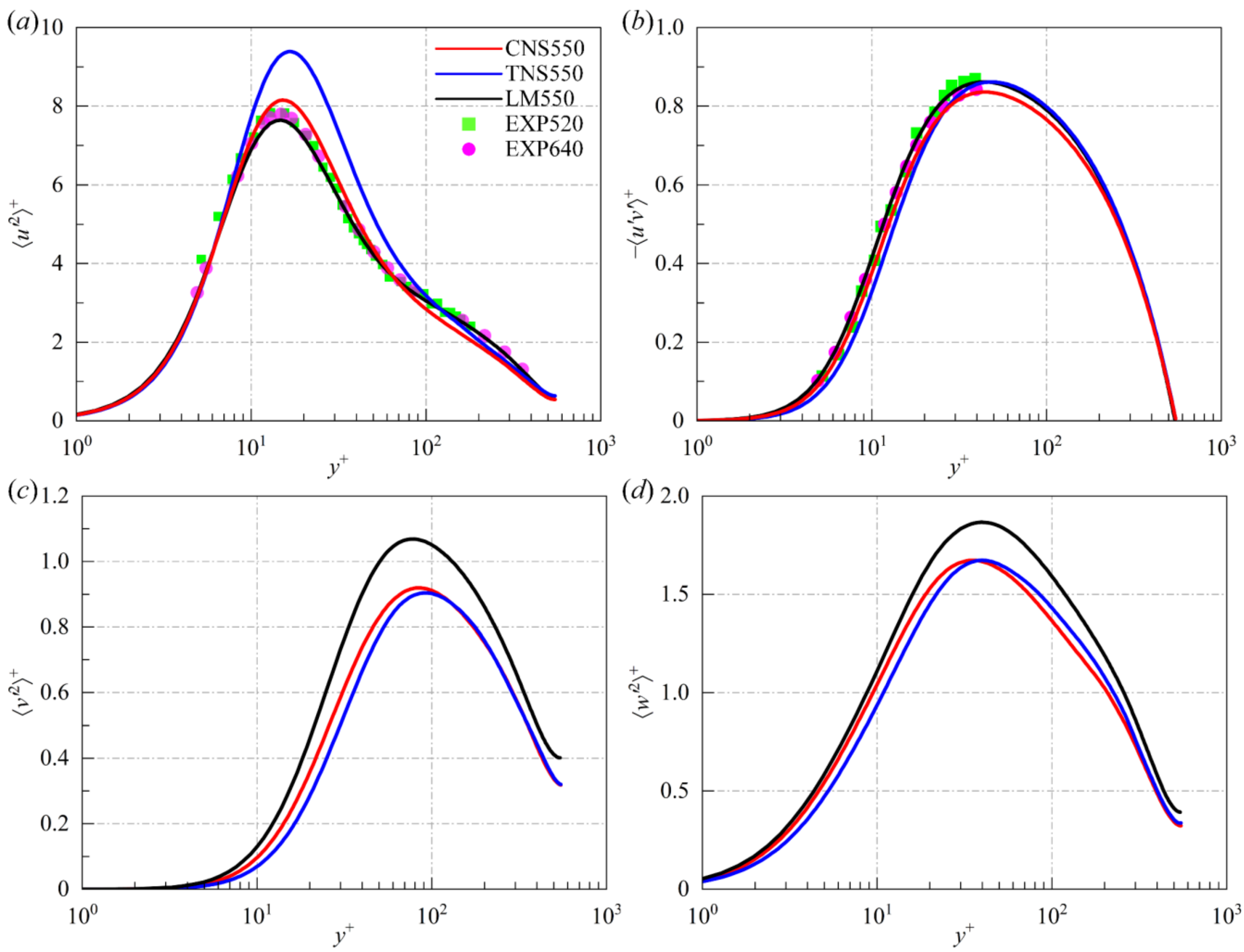

FIG. 5. Profiles of the (a) streamwise Reynolds normal stress $\langle u'^2 \rangle^+$, (b) Reynolds shear stress $-\langle u'v' \rangle^+$, (c) wall-normal Reynolds normal stress $\langle v'^2 \rangle^+$ and (d) spanwise Reynolds normal stress $\langle w'^2 \rangle^+$ with the lines and symbols being defined in Table III.

Similar discrepancies are observed in the wall-normal $\langle v'^2 \rangle^+$ and spanwise $\langle w'^2 \rangle^+$ Reynolds normal stresses, as given in Figs. 5(c) and (d). In the TNS, the energy transfer from the streamwise component to the transverse directions is regulated by both pressure-strain correlations and viscous dissipation (Mansour et al. 1988). In the CNS, however, removing the viscous resistance to stretching deformation permits the relatively unimpeded stretching of fluid elements in the wall-normal and spanwise directions, which enhances the efficiency of the pressure-strain term in redistributing energy from the streamwise component to the transverse fluctuations, as evidenced by the inner peaks predicted by the CNS exceeding the TNS data by 2% in Figs. 5(c) and (d). The peak locations for these transverse components also shift wall-ward by 13%, which further confirms that the turbulent structures in the CNS are spatially compressed toward the wall, resulting in a thinner viscous sublayer and consequently elevating near-wall friction.

Figure 6 presents the profiles of the pressure fluctuation intensity, $p'^{+}$, across all the cases. The CNS predictions are consistently higher than the TNS baseline and approach the LM data. Specifically, the peak magnitude of the CNS is about 3% higher than that of the TNS, with the elevation being more pronounced in the vicinity of the wall than in the core flow region. The phenomenon is primarily attributed to the enhanced wall friction predicted by the CNS, which significantly amplifies the mean velocity shear near the wall. Consequently, the rapid pressure term in the Poisson equation is intensified by the dominant mean-shear/turbulence interactions. Simultaneously, the CNS shifts the $y^{+}$ location of the pressure peak closer to the wall than TNS by about 8%. The spatial shift results from the absence of the stretching viscous dissipation, which permits the turbulent coherent structures being further compressed toward the near-wall region, giving rise to the thinner viscous sublayer, and thereby driving the physical loci of the maximum turbulence production and pressure fluctuations toward the solid boundary.

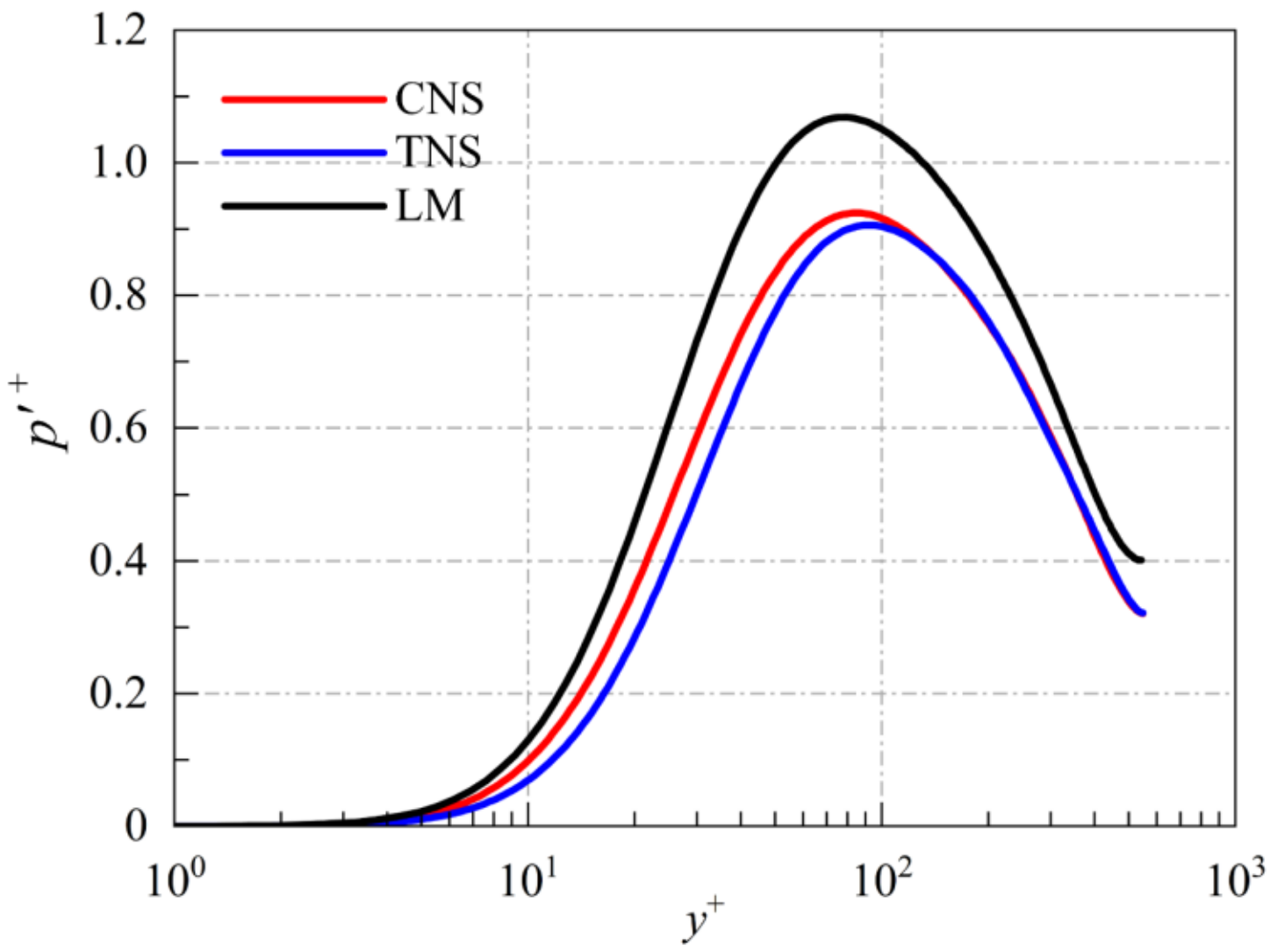

FIG. 6. Profiles of the total pressure intensity $p'^{+}$ with the CNS, TNS, and LM results being denoted by the red, blue, and black lines, respectively.

To further explicitly interrogate the effects of eliminating the stretching stress tensor, the component profiles in the added stress tensor of $\boldsymbol{F}^{add}$ are presented in Fig. 7. Figure 7(a) displays the streamwise normal component of $\left\langle F_{xx}^{add} \right\rangle^{+}$ featuring with the distinctive double-peak structures, namely an inner peak around $y^{+} \approx 5-10$ corresponding to the streak generation in the buffer layer, and an outer peak ($y^{+} \approx 60-70$) reflecting the modulation of the near-wall region by large-scale motions (LSMs). Figure 7(b) depicts the wall-normal component, $\left\langle F_{yy}^{add} \right\rangle^{+}$, which exhibits the clear sign of reversal. The positive values near the wall correspond to the 'splatting' effect (Perot & Moin 1995), namely the rapid deceleration of fluid elements impinging onto the solid boundary, while the negative values in the outer region are associated with the local deformations during the ejection ($v' > 0$) and sweep ($v' < 0$) events in turbulence, typically the hairpin vortices, as detected in Figs. 10 and 11 by the Rortex iso-surfaces. The spanwise component, $\left\langle F_{zz}^{add} \right\rangle^{+}$, as shown in Fig. 7(c), exhibits the same trend as the streamwise component. The similarity arises because the streamwise streak formation is intrinsically coupled with the spanwise deformation, which is a critical factor in the self-sustaining mechanism of wall turbulence (Hamilton et al. 1995). Furthermore, Fig. 7(d) presents the shear component, $-\left\langle F_{xy}^{add} \right\rangle^+$, which features a characteristic single peak within the buffer layer ($y^+ \approx 5 - 10$), precisely coinciding with the region of maximum turbulence production. The shear component, as an important driving force, participates in the process of regulating the momentum balance in Eq. 3.5, resulting in a decrease of the Reynolds shear stress, $-\langle u'v' \rangle^+$, in the CNS as seen in Fig. 5(b).

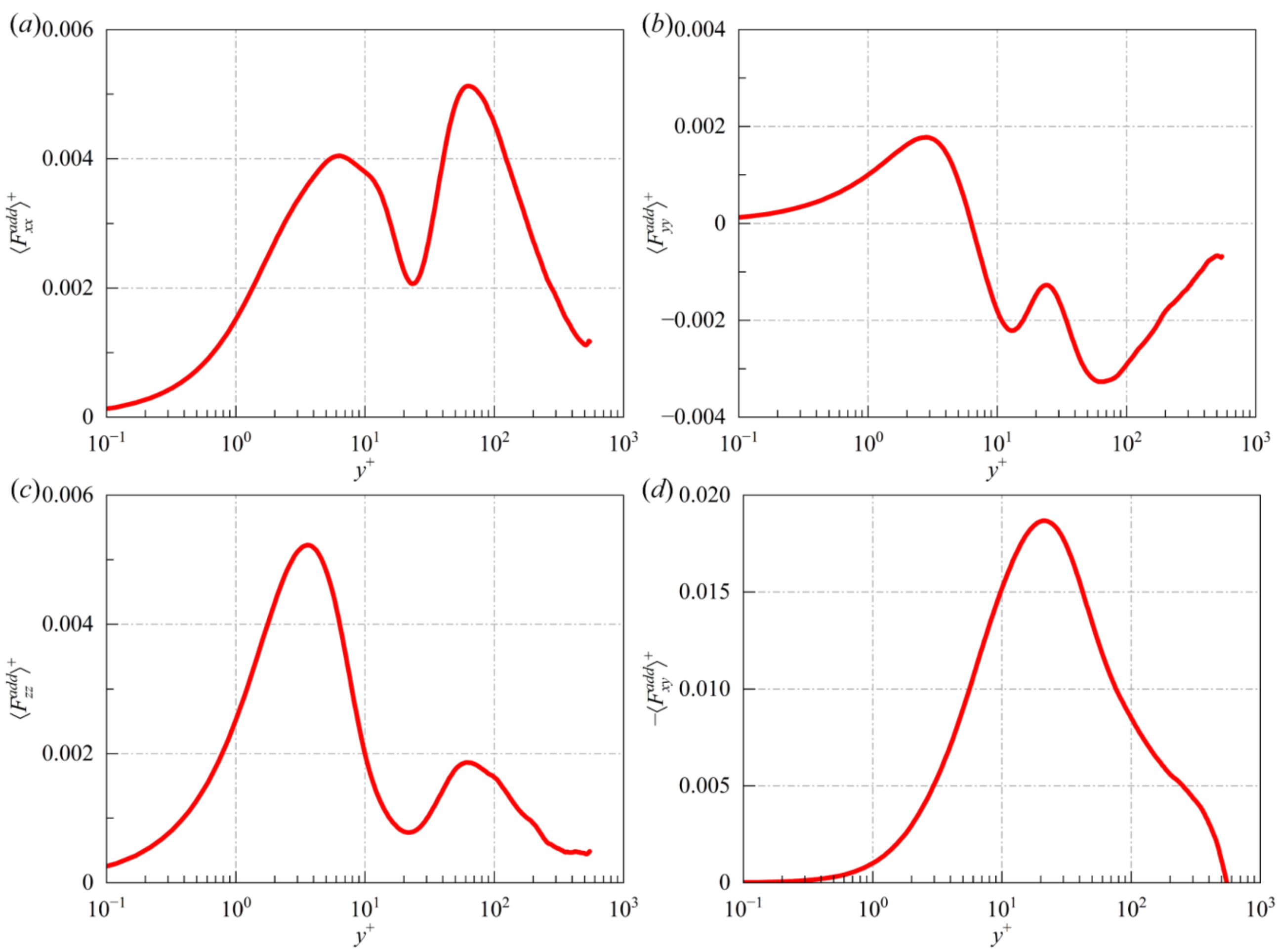

FIG. 7. Profiles of the major components in the added stress tensor $\boldsymbol{F}^{add}$: (a) $\left\langle F_{xx}^{add} \right\rangle^+$; (b) $\left\langle F_{yy}^{add} \right\rangle^+$; (*c*) $\left\langle F_{zz}^{add} \right\rangle^+$; (*d*) $-\left\langle F_{xy}^{add} \right\rangle^+$.

*4.1.3 TKE budgets and energy spectral density*

The budget equation for the turbulent kinetic energy (TKE) is described by Mansour et al. (1988) and Hoyas & Jiménez (2008) as

$$B = P_K + \varepsilon + T + \Pi^s + \Pi^d + V. \tag{4.5}$$

The terms on the right-hand side stand for the production, dissipation, turbulent diffusion, viscous diffusion, pressure-strain, and pressure-diffusion, respectively. Figure 8 illustrates the profiles of the production ($P_K$), dissipation ($\varepsilon$), and the relative imbalance terms. The production term, $P_K$, displays a distinct single peak in Fig. 8(a). The peak magnitude from the CNS is significantly higher than that of the TNS and shifts closer towards the wall by penetrating deeper into the buffer layer. Regarding the dissipation of $\varepsilon$ with its maximum at the wall, the CNS prediction substantially exceeds the one from TNS throughout the near-wall and buffer regions, with the maximum discrepancy occurring at the wall in Fig. 8(b). Furthermore, Hinze (1975) posited the existence of an intermediate region between the inner and outer layers where the transport terms were negligible compared to production, yielding $P_K \approx \varepsilon$. The relative imbalance between production and dissipation ($P_K / \varepsilon - 1.0$) is plotted in Fig. 8(c). The TNS profile clearly exhibits a stable plateau where the imbalance approaches zero. In contrast, although the CNS profile also displays a relatively broader plateau, an imbalance can reach approximately 20% and the increase is evidently due to the relative enhancement of the dissipation rate in CNS, which exceeds the production term in TNS. The observation suggests that in the traditional Newtonian constitutive relation, the stretching deformation essentially acts as a 'damper' and 'buffer' for the energy cascade in the near-wall region, suppressing the excessive local energy production and dissipation. The CNS breaks the constraint, resulting in a more intensified and highly coupled cycle of TKE production and dissipation.

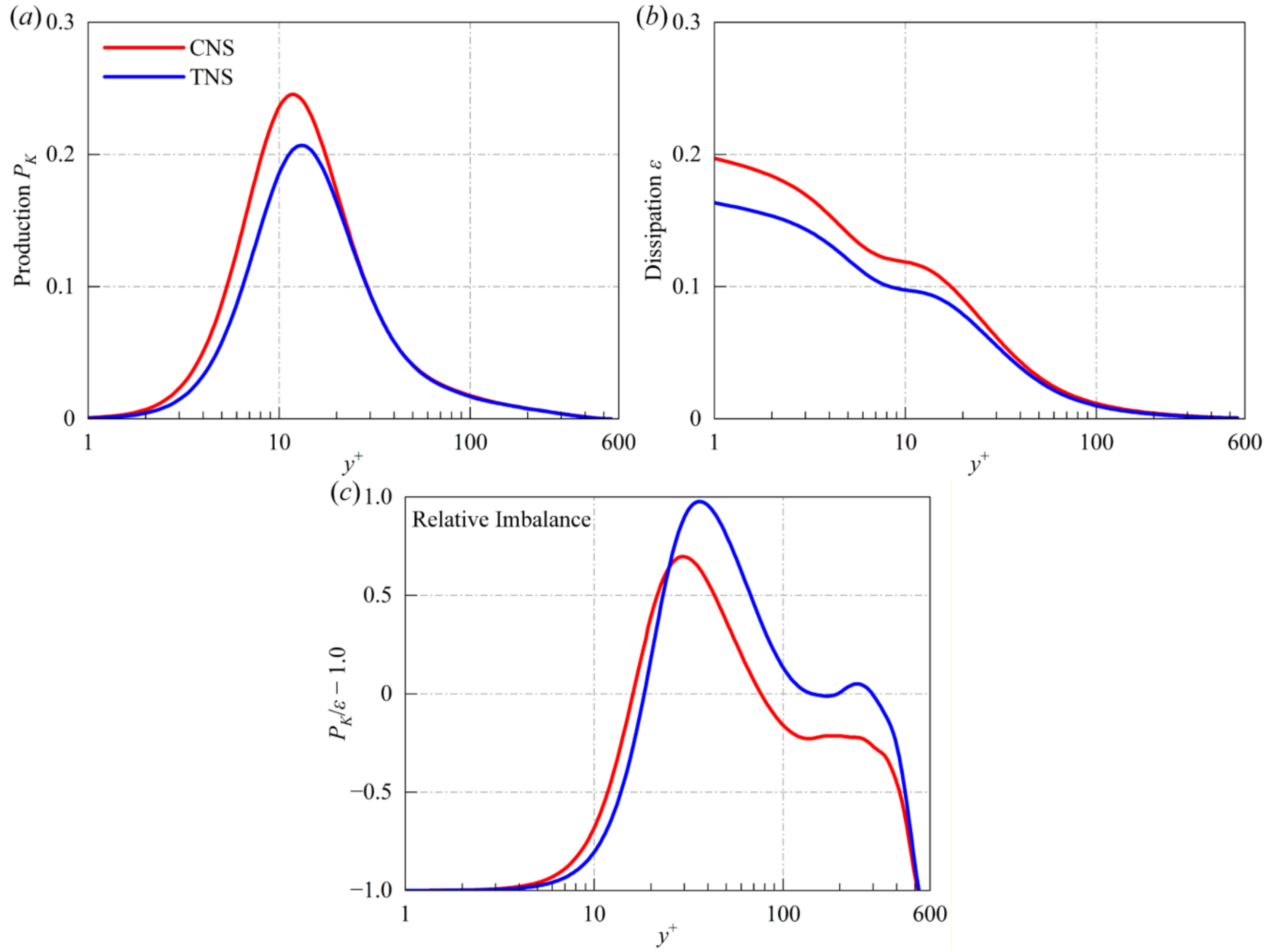

FIG. 8. Turbulent kinetic energy budgets: (a) production $P_K$, (b) dissipation $\varepsilon$, and (c) relative imbalance terms with the CNS and TNS results being denoted by the red and blue lines, respectively.

Figure 9 presents the premultiplied spectral energy density of the streamwise velocity fluctuations. Because the premultiplied spectrum represents the energy density per unit logarithm of the wavenumber, the logarithmic scale employed here effectively illustrates the energy distribution across various scales. The high-energy core region and its surrounding contours in the CNS exhibit a significantly broader distribution than that in TNS, which evidently suggests that the elimination of stretching deformation substantially alters the energy cascade and distribution pathways in the channel turbulence, weakening the energy convergence at a single characteristic scale and causing the TKE to disperse and to reorganize over a broader range of scales and wall-normal space. Simultaneously, the core location of the CNS spectrum shifts directionally, not only moving closer to the wall (lower $y^+$) but also shifting toward longer streamwise wavelengths, which is confirmed being consistent with the results of the spatial structures of the SPOD modes of vortices in Fig. 14. The observation is attributed to eliminating stretching stress in CNS, which is demonstrated acting as a form of viscous damping in the TKE budgets of Fig. 8, and as a result, the coherent structures undergo the less constrained elongation.

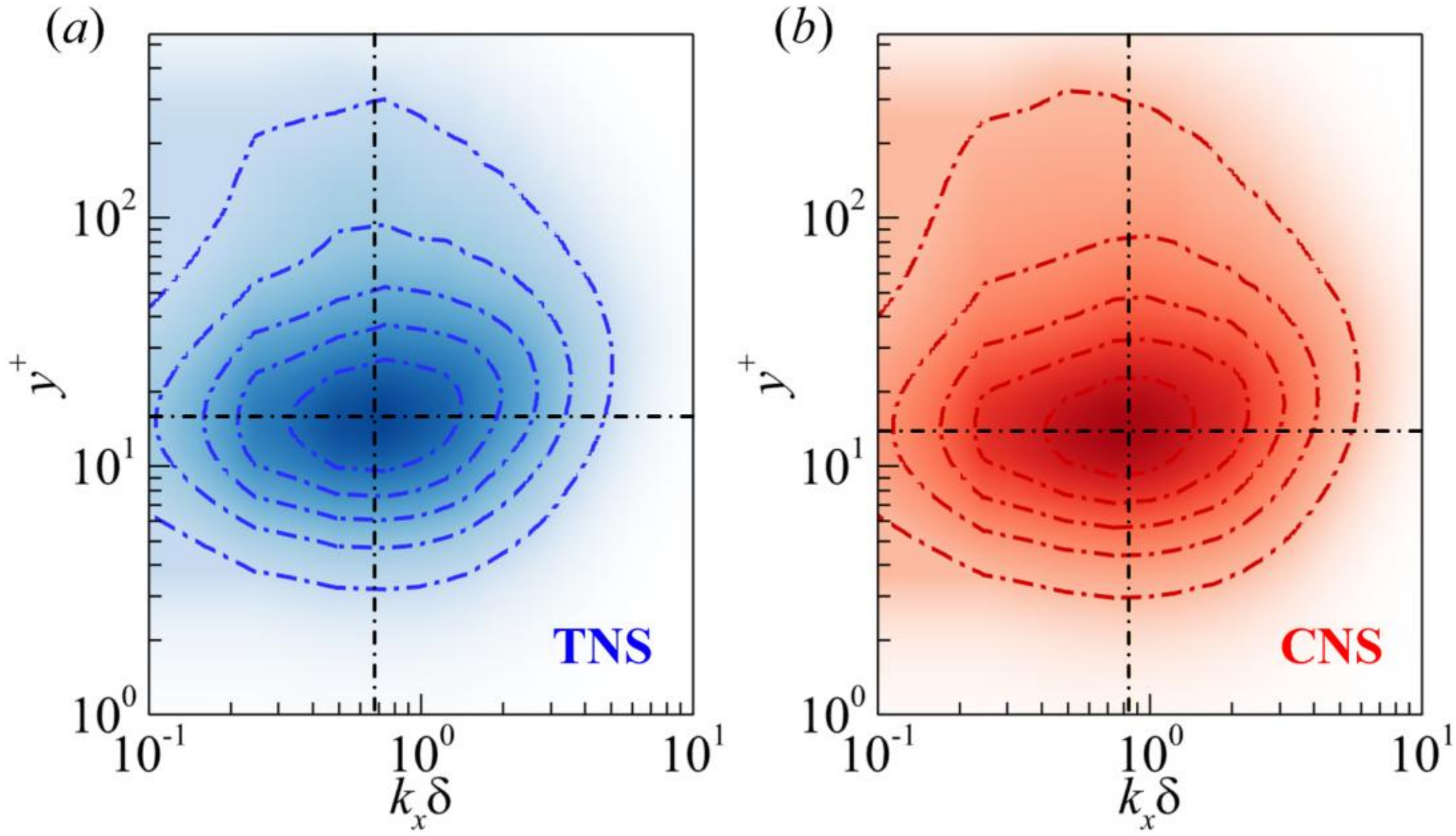

FIG. 9. Wavenumber-premultiplied energy spectral density $k_xE_{uu}/u_\tau^2$ for (a) TNS and (b) CNS.

### 4.2 Instantaneous coherent structures of channel turbulence

The Rortex quantity and the associated R-R-S decomposition offer the advantage of effectively decoupling the shear contamination in TBL and accurately capturing the rigid rotation of fluid elements. The CNS equations further push the frontier of fluid kinematics based on Rortex into fluid dynamics, which provides an opportunity to more precisely bring out the fundamental and complex morphology of vortical structures in turbulence.

Figure 10 visualizes the instantaneous coherent structures contained in the DNS data of channel turbulence generated by both TNS and CNS and identified via the iso-surfaces of Rortex magnitude with a threshold of $|\boldsymbol{R}| = 15$. From the visualization, the fully developed turbulence in channel is observed being predominantly populated by the near-wall streaks, presenting as the mixed vortical structures with the distinctive quasi-streamwise and hairpin patterns. By comparing the data at the $|\boldsymbol{R}| = 15$ iso-surfaces, both the TNS and CNS bring out the canonical structures near wall as highlighted in Fig. 10 by the pink boxes with the arch-hairpin shape and typical streaky pattern. Through a more detailed comparison of these individual structures, the arch hairpins and the streaks in the CNS tend to be qualitatively thinner and smaller than those in the TNS, which motivates the subsequent SPOD analysis to more quantitatively uncover these structures in terms of their spatiotemporal characteristics in both modes and frequencies. These observations in Fig. 10 evidently suggest that the near-wall turbulence regeneration cycle is fundamentally sustained by the shearing-only motions in TBL and the stretching-induced viscous stresses in TNS do affect the morphology features of coherent structures. Within the context, the threshold-free vortex identification (VI) approach is further explored in the current study to more physics-aligned and mathematics-rigorously defined the vortex structures.

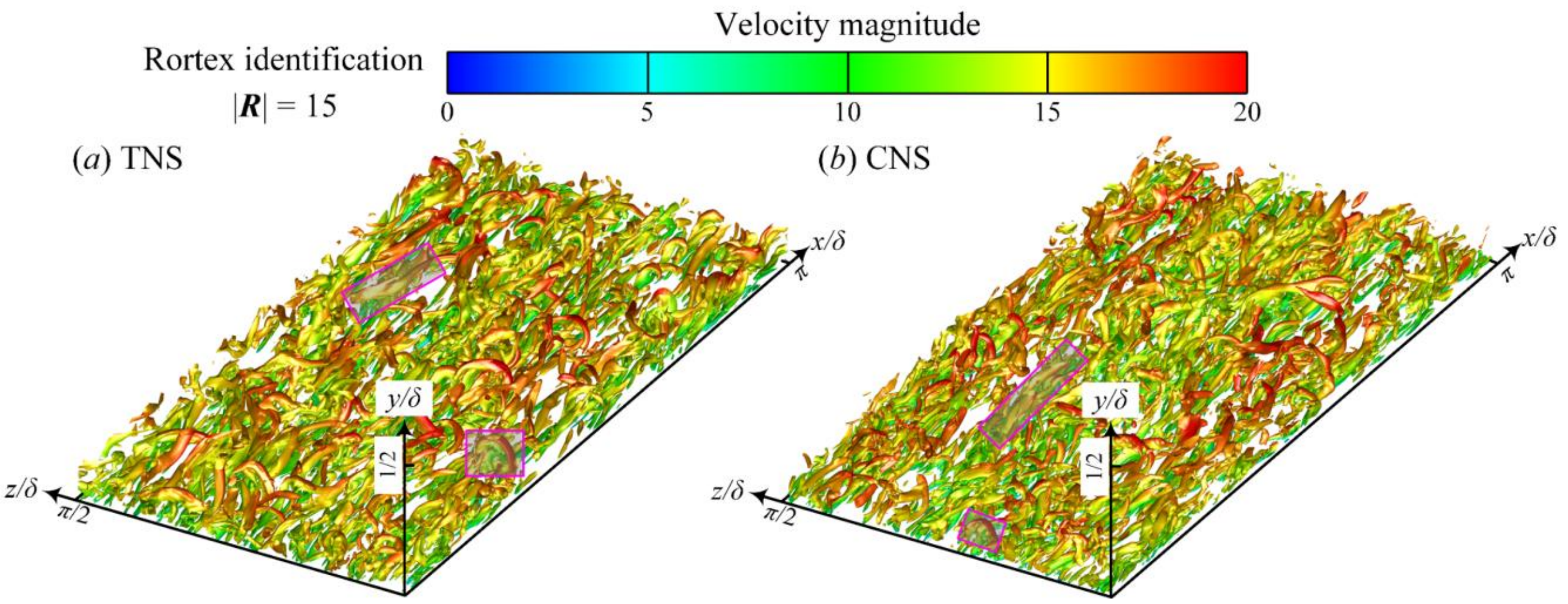

FIG. 10. Instantaneous vortical structures visualized by the $|\boldsymbol{R}| = 15$ iso-surfaces for (a) TNS and (b)

CNS, colored by velocity magnitude.

Based on the discriminant $\Delta$ of the eigen-equation for VGT, a more rigorous kinematic classification can be defined in a flow field, as commented in **Section 2.3**. Figure 11 illustrates the spatial topological distribution of the rotational regions at an arbitrary time instant as demarcated by the iso-surfaces of $\Delta$ in a fully developed turbulent channel flow. The red regions ($\Delta > 0$) characterize the rotation-dominated zones, the blue-colored or hollow 3D space standing for the non-rotational or shearing-only region ($\Delta < 0$), and the green regions ($\Delta = 0$) precisely delineate the RNRI proposed in the present study. As depicted, the RNRI in the fully developed turbulent field exhibits a highly dense and spatially interweaving iso-surface topology, which serves as the bounding surface tightly enveloping the local rotational domains, and thereby intuitively exposes the distinctive rotating and shearing regimes ubiquitously contained in their nonlinear dynamic evolution in turbulence.

In the 3D perspective as shown in Figs. 11(c) and (d), the green RNRI structures in the near-wall region manifest as anisotropic, streamwise-elongated tubular structures corresponding to near-wall streamwise streaks, which gradually evolve into the hairpin-patterned iso-surface structures in the logarithmic and outer regions. In the streamwise–wall-normal ($x - y$) plane as shown in Figs. 11(a) and (b), the red rotating regions and the blue non-rotating regions exhibit an alternating distribution. The red vortex cores display an inclined lifting motion, reflecting the upward ejection ($Q_2$) of low-speed fluid and the wall-ward sweep ($Q_4$) of high-speed fluid. Furthermore, the correction introduced by the CNS over the TNS manifests as more fragmented vortical structures and more corrugated RNRI patterns, indicating an intensified interaction between the rotating and shearing regions.

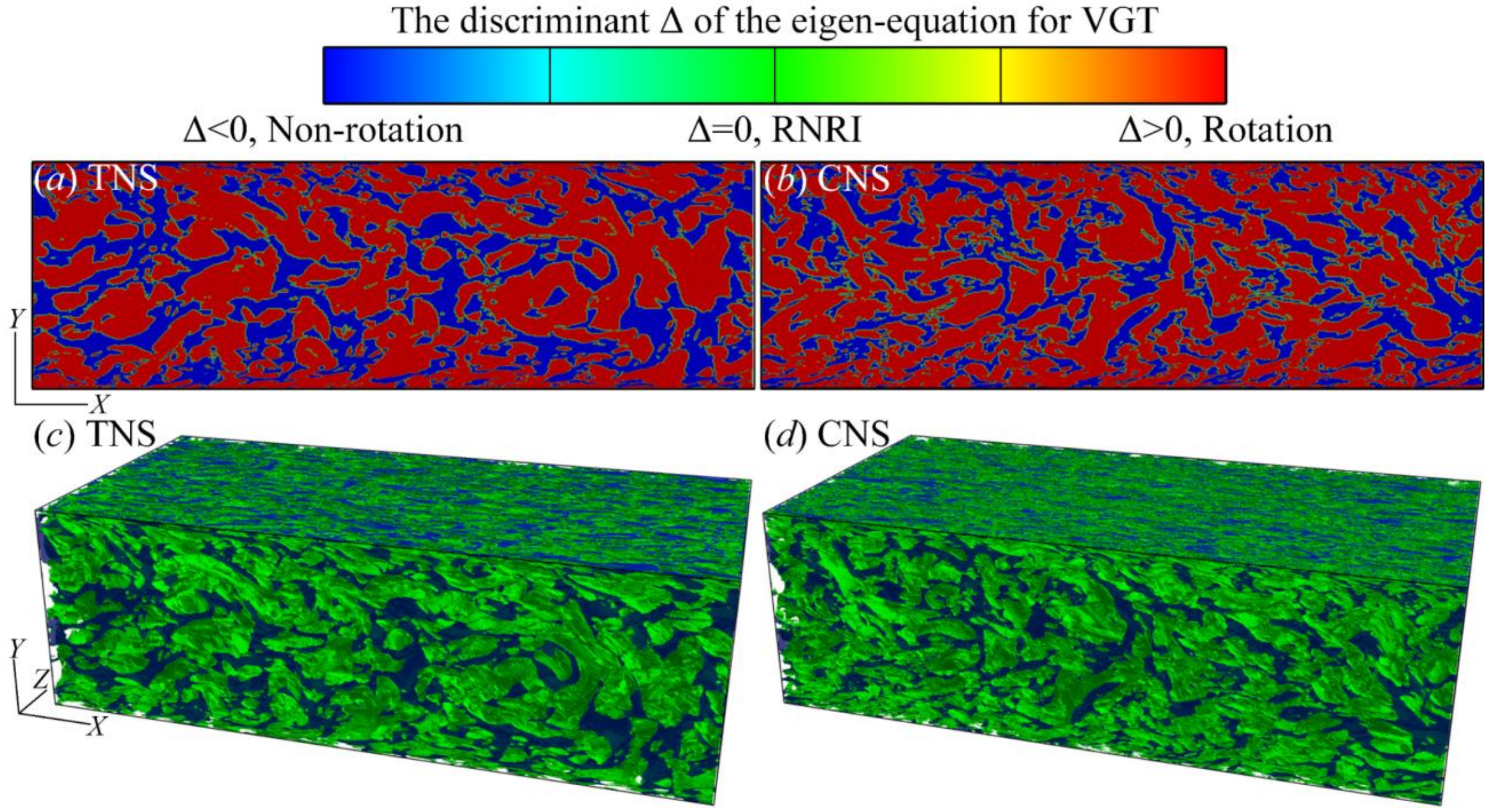

FIG. 11. Schematic of fully developed channel turbulence with the green iso-surfaces representing the RNRI ($\Delta = 0$), the blue-colored or hollow 3D space standing for the shearing-only region ($\Delta < 0$) and the red-colored space marking the rotational region ($\Delta > 0$).

Given the highly dense and complex interwoven nature of the vortex-cell structures depicted by RNRI in fully developed channel turbulence, Fig. 12 utilizes the laminar-turbulent transition in the flat-plate boundary layer to more clearly and conceptually delineate the hierarchy of large-scale vortex-cell characteristics. The figure comprehensively brings out the space-sequential evolution of these vortex events in the TBL natural transition, namely the large-scale vortex-cell structures progressively developing from the Tollmien-Schlichting (T-S) waves, to the three-dimensional distortion of Λ-shape vortices, and ultimately to the incipient breakdown of complex hairpin vortices.

Through applying the iso-surface visualization to the RNRI ($\Delta = 0$) instead of a traditional VI quantity ($Q$, $\lambda_{ci}$ or even $|\boldsymbol{R}|$), the 3D view in Fig. 12(a) and the top view in Fig. 12(b) meticulously anatomize the geometric topologies nesting within their specific cell spaces. The unique representation method comprehensively delineates each fluid vortex-cell structure without resorting to any arbitrary threshold. From the exterior to the interior, the cell structure contains the rotational/non-rotational interface (RNRI, $\Delta = 0$, 2D) playing as the vortex-cell boundary, the rotational domain (RO, $\Delta > 0$, 3D) of a 3D vortical space, and the Rortex core line (CL, $\nabla \boldsymbol{R} \times \boldsymbol{R} = 0$, 1D) representing the vortex core or the cell-skeleton with the central axis of rotation. Regarding the conventional vortex structures based on the iso-surface of a VI scalar, Fig. 12 also provides an iso-surface of $|\boldsymbol{R}| = 0.01$, which is an entity enwrapped inside the RNRI boundaries.

Regarding the specific graphical representation, the 3D view in Fig. 12(a) presents the Rortex core lines (the black lines) constituting the topological skeletons in the aforementioned transition stages. Through demonstrating the Rortex vector $\boldsymbol{R}$ and the vorticity vector $\boldsymbol{\Omega}$ at a selected point in the rotational domain, it can be clearly seen that $\boldsymbol{R}$ aligns exceptionally well with the direction of the vortex core line. Meanwhile, $\boldsymbol{\Omega}$ points in an almost perpendicular or an irrelevant direction to the core line, which totally excludes the possibility of applying the conventional vorticity as the VI quantity.

The partition of the flow field based on VGT eigen status possesses its profound physical implications. To examine the physical interpretation of the 2R situation or RNRI, a local eigen decomposition is performed in Fig. 12. The eigenvector $\boldsymbol{e}_1$, corresponding to the independent real root, characterizes the principal stretching trend of fluid elements on the RNRI surface, which serves as a continuous topological extension of the internal vortex core direction onto its outer boundary. Conversely, the generalized eigenvectors $\boldsymbol{e}_{2,3}$, associated with the double root, directly represent the collapsing asymptotes of the rotational plane. These vectors intersect non-orthogonally with $\boldsymbol{e}_1$, exposing the fluid shear-slip mechanism. Combined with the top view, it can be intuitively observed that as the instability process develops, the RNRI concurrently undergoes complex wrinkling and reconnecting events. (The zig-zags on the RNRI surfaces are obviously due to the grid resolution issue, particularly for the more refined vortex structure entities).

Therefore, the hierarchical structure (RNRI−RO−CL) is quite well validated for its mathematical rigor and physical soundness in turbulence research, particularly with the $\Delta = 0$ being found as the natural demarcating boundary for rotation and non-rotation or as the border of each vortex-cell. Consequently, the innovative VI method based on the RNRI and Rortex core-line provides a valid and unique approach from a fluid kinematic perspective for comprehensively and accurately characterizing a vortex-cell structure with the evident mathematical (rigorous) and physical (soundness) advantages in studying the underlying dynamics in turbulence.

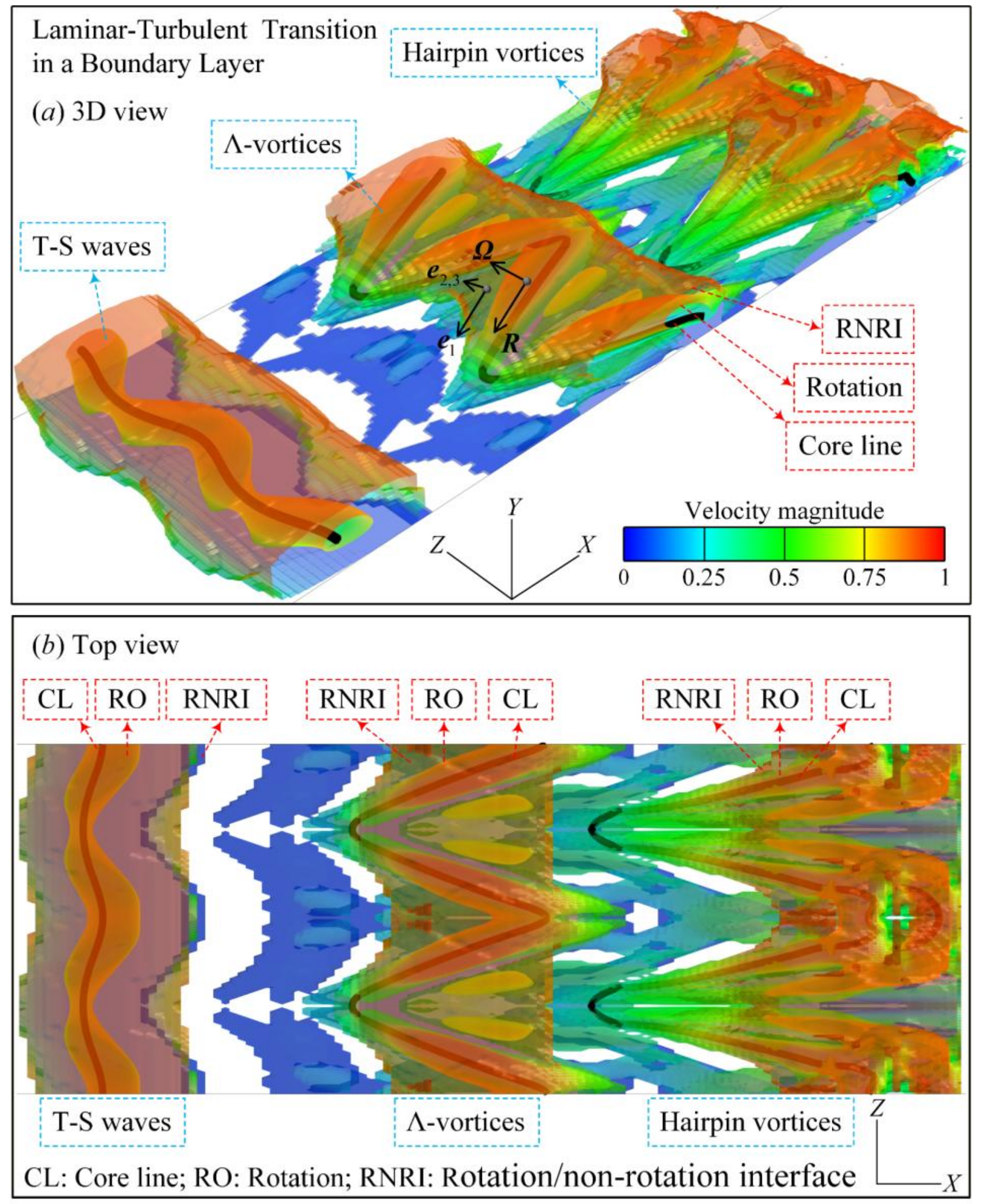

FIG. 12. Schematic of laminar-turbulent transition in a flat-plate boundary layer: (a) 3D view; (b) top view, with the vortex-cell components being outlined successively from the outer border to the inner core, namely the (1) RNRI ($\Delta = 0$), (2) rotation ($\Delta > 0$) with the iso-surface of $|\boldsymbol{R}| = 0.01$, and (3) Rortex core line with the axis of rotation.

### 4.3 SPOD analysis of channel turbulence

In the current section, the reduced-order decomposition and frequency-spectral analysis are provided based on performing the SPOD of DNS data on the streamwise and wall-normal plane ($x - y$) in the turbulent channel flow at $Re_\tau = 550$ generated by both the TNS and CNS models. In all the studied cases, the ensemble sampling number was selected at $n_t = 3000$ snapshots. The data were segmented into $n_{blk} = 22$ blocks with $n_{fft} = 256$ snapshots per block and 50% overlap.

Since SPOD yields one set of eigenpairs per frequency, the contributions from individual frequency can be investigated independently. The resulting SPOD eigenvalues form a continuous energy spectrum, as depicted in Figs. 13(a) and (b) for the pressure field $p$. The first mode given by the blue line represents the most energetic structures and is henceforth referred to as the leading mode. Given that the first two modes capture a vast majority of the turbulence energy (approximately 90%), the characteristic frequencies are then extracted by focusing on the two modes.

It is noteworthy that Fig. 13 displays the noticeable sinusoidal spectra in the low frequency regime, with the multiple energy peaks, which is referred to as the convective nonlinearity phenomenon with its mechanism being satisfactorily explained by the nonlinear convection terms in both TNS and CNS. Due to the velocity product form of convection terms, the higher harmonics emerge as the integer multiples of a baseline frequency. Therefore, via denoting the $i$-th characteristic frequency as $St_i$ and taking $St_1$ as the baseline frequency, it is intriguing to observe that the subsequent peaks occur at $St_2 \approx 2St_1$ and $St_3 \approx 3St_1$, which suggests that the higher-order characteristic frequencies are harmonics of $St_1$. The progression from the baseline frequency to higher harmonics forms an energy cascading process from large to small scales mainly driven by the convective terms in both TNS and CNS, which is here well proved and identified by the SPOD spectra in Fig. 13. Moreover, $St_1$ is physically found corresponding to structures with a streamwise wavelength of $\lambda \approx 10\delta$ associated with the very-large-scale motions (VLSMs; del Álamo et al. 2004). The third harmonic of $St_3$ is then linked to $\lambda \approx 3 \sim 4\delta$, representing the hairpin vortex packets (Hwang & Cossu 2010), whereas the second harmonic of $St_2$ demonstrates its connection to $\lambda \approx 5 \sim 6\delta$, reflecting the spatial amplitude modulation of these hairpin packets (Mathis et al. 2009). All these evidences suggest that the convective nonlinearity phenomenon applies not only in temporal but also in spatial direction, which is then elucidated quantitatively in the following SPOD mode study. Consequently, through the SPOD reduced-order decomposition, the complex multi-scale turbulence is decoupled into canonical structures collectively characterized by these individual scales and frequencies, thereby successfully reconstructing and uncovering the large-to-small-scale energy cascading pathways.

Furthermore, a regular frequency shift can be identified in Fig. 13(c), when comparing the spectra from the CNS and TNS models, which can obviously be attributed to the correction terms of $\boldsymbol{F}^{add}$ in CNS. Specifically, the odd-order frequencies decrease (e.g., $St_1$ shifts from 1.52 to 1.41, and $St_3$ from 4.53 to 4.41), while the even-order frequencies increase (e.g., $St_2$ increases from 2.89 to 2.96). Given the correspondence between the frequency and wavelength, the shift in odd-order frequencies implies a longer spatial wavelength for the vortex packets, whereas the shift in even-order frequencies indicates a denser spacing between vortices. The behavior is consistently corroborated by the spatial structures of the modes, as depicted in Fig. 14.

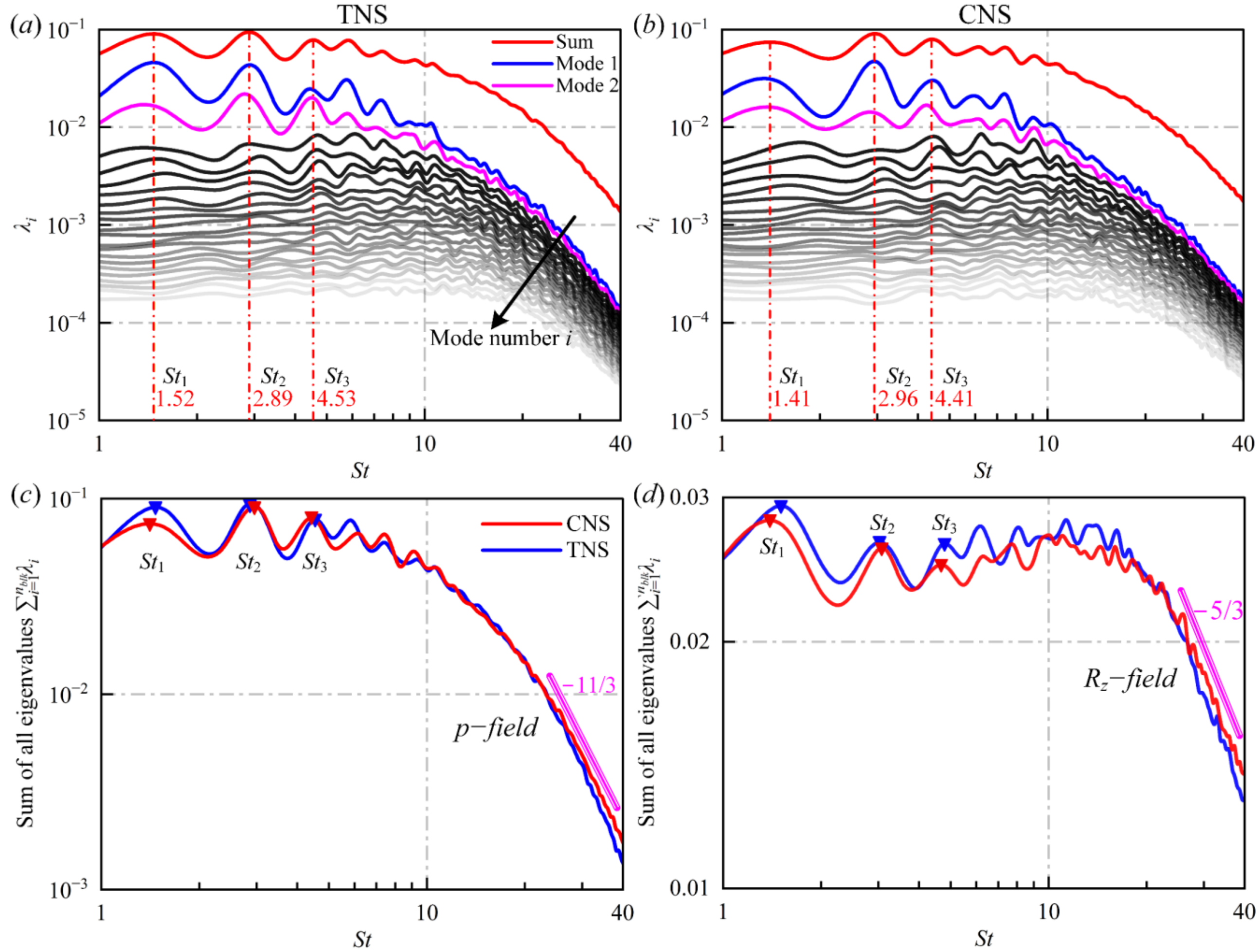

FIG. 13. SPOD energy spectra in channel turbulence. Modal energy spectra of $p$ for the (a) TNS and (b) CNS models, with red, blue, pink, and black lines denoting the sum of all eigenvalues, mode 1, mode 2, and higher modes, respectively, and $St_i$ representing the characteristic frequency of the $i$-th energy peak. Total energy spectra of (c) $p$ and (d) $R_z$, with red and blue lines representing the CNS and TNS models, respectively.

Figure 13(c) presents the total SPOD energy spectra of the $p$ for the TNS and CNS models. The energy spectra for both models exhibit an approximate −11/3 decay rate (Bradshaw 1967; Panton & Linebarger 1974; George et al. 1984)], which is attributed to the pressure fluctuations dominated by the mean shear-turbulence interactions. In a wall-bounded turbulence with strong mean shear, such as the channel turbulence, these fluctuations arise primarily from the coupling between the mean shear and turbulent velocity fluctuations. Within the high-frequency small-scale regime of $St > St_2$, the turbulence rapidly enters the dissipation range, where the energy decays exponentially owing to the viscous effects. Because the CNS eliminates the dissipation pathway associated with the stretching deformation, the increased energy cascading from the enhanced large-scale production cannot be efficiently dissipated through the normal straining mechanisms. Consequently, the turbulent energy piles up at the small scales, leading to a more gradual high-frequency spectral attenuation in CNS relative to TNS. Conversely, in the low-frequency regime of $St < St_2$, which corresponds to the large-scale coherent structures, the elimination of stretching deformation in the CNS slightly adjusts their self-sustaining processes. The interference marginally impairs the structural coherence and energy-containing capacity of the large-scale eddies, yielding a slightly attenuated low-frequency energy peaks in CNS data.

The total SPOD energy spectra are further presented for the spanwise Rortex ($R_z$) component in Fig. 13(d), which brings out a broad energetic plateau in the range of $St < 15$ and a distinct power-law decay for $St > 15$. As a result, the dissipative damping is significantly attenuated for the high-frequency ($St > 15$) small-scale vortices. Similarly, due to the accumulation of energy on a small scale, the high-frequency ($St > 15$) tail energies of CNS are consistently higher than those in TNS, and exhibit a slower decay rate. Furthermore, the high-frequency decay rates of $R_z$ spectrum of both TNS and CNS, characterizing the rigid-rotation intensity, agree well with the classical Kolmogorov −5/3 scaling law, which was corroborated by Xu et al. (2019) and was further confirmed by the current SPOD spatiotemporal spectra.

Figure 14 provides the spatial structures of the first two modes of $p$ on the streamwise and wall-normal plane ($x - y$), where $St_iM_j$ represents the $j$-th Mode at the $i$-th characteristic frequency. Across all the cases, the pressure modes manifest as the alternating positive and negative wave packets generally elongated in the streamwise direction. The spatial distributions exhibit a pronounced forward inclination, reflecting the convective nature of coherent structures in a typical wall-bounded turbulence. Specifically, the modal spatial structure at the low-frequency $St_1$ corresponds to the large-scale streaks, $St_3$ represents as hairpin vortex packets, and $St_2$ characterizes the inter-scale modulation.

The streamwise projected distance between the extrema of two adjacent vortex packet cores is defined and denoted by $\xi_{i,j}^+$ as the characteristic wave-packet length for the $j$-th mode at the $i$-th characteristic frequency. Thereby, Table V summarizes these characteristic wave-packet lengths across all the cases in the study. Analogous to the temporal harmonics observed in the characteristic frequencies, the characteristic wave-packet lengths manifest a spatial harmonic relationship, namely $\xi_{2,1}^+ \approx 2\xi_{3,1}^+$ and $\xi_{1,1}^+ \approx 3\xi_{3,1}^+$. This spatiotemporal harmonic phenomenon jointly confirms that for the vortical structures (e.g., $St_1$ and $St_3$), the CNS yields a longer streamwise wavelength and a smaller vortex spacing (due to the modulation by $St_2$) comparing to those in TNS model. Furthermore, the CNS demonstrates a more pronounced stratification in the wall-normal direction, manifesting as an increased wall-normal wavenumber, along with a larger tilted angle of vortex packets in Fig. 14. Finally, the leading mode ($M_1$) and the suboptimal mode ($M_2$) yield similar and consistent conclusions, whilst their distinct scales collectively capture the multi-scale characteristics inherent in channel turbulence.

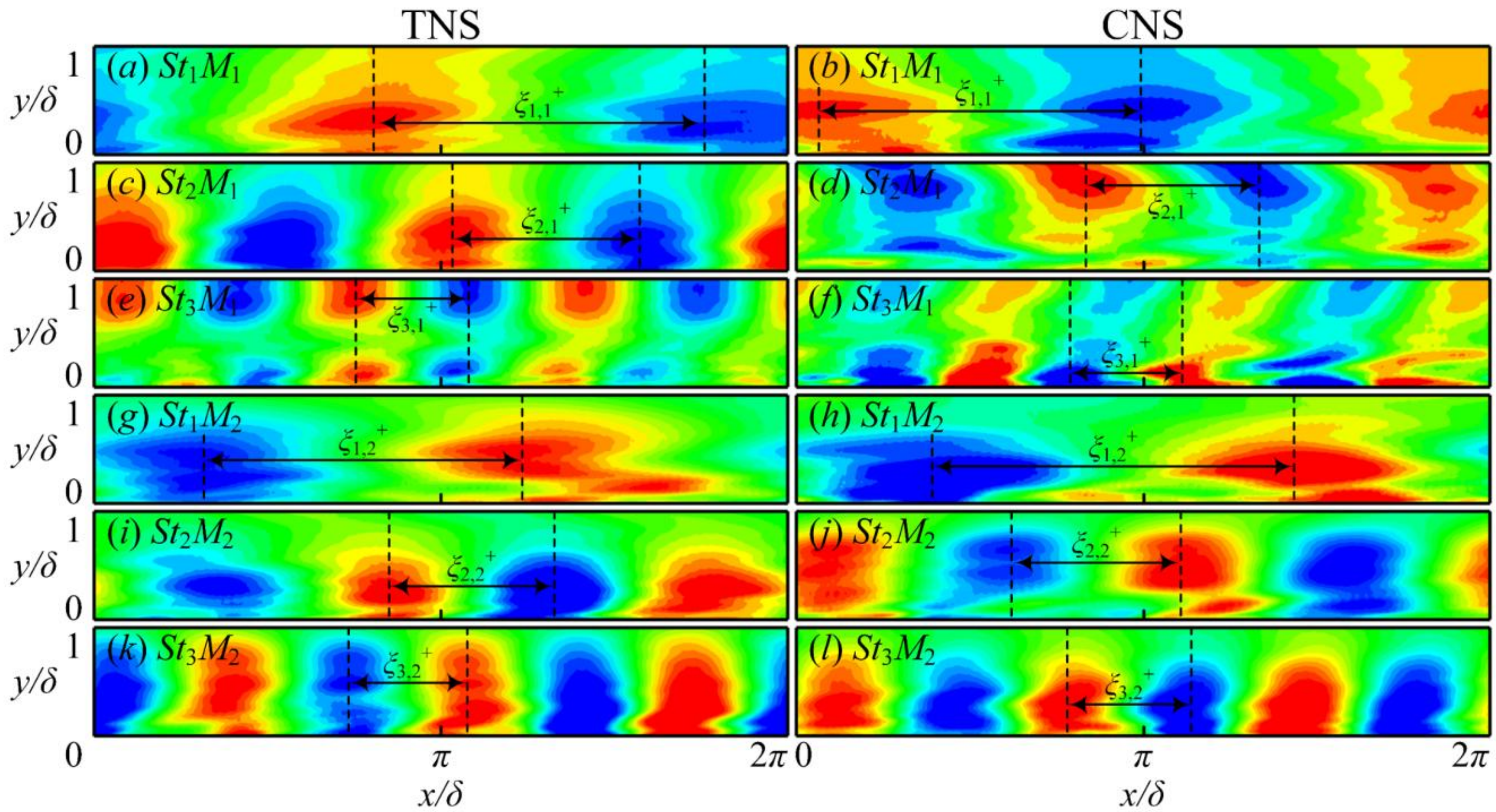

FIG. 14. Spatial structures of the SPOD modes of $p$ for TNS and CNS models, where $St_iM_j$ represents the $j$-th mode at the $i$-th characteristic frequency and $\xi$ is defined as the characteristic wave-packet length in the streamwise direction.

TABLE V. Characteristic wave-packet lengths, $\xi_{i,j}^+$, of the SPOD pressure modes for the TNS and CNS models, corresponding to the $j$-th mode at the $i$-th characteristic frequency.

| Model | $\xi_{1,1}^+$ | $\xi_{2,1}^+$ | $\xi_{3,1}^+$ | $\xi_{1,2}^+$ | $\xi_{2,2}^+$ | $\xi_{3,2}^+$ |
|---|---|---|---|---|---|---|
| TNS | 1605 | 926 | 557 | 1578 | 817 | 585 |
| CNS | 1610 | 857 | 567 | 1789 | 807 | 618 |

Given that the SPOD modes are spatially coherent and contain the most energy at each frequency, a SPOD-based frequency−time analysis can be used to gauge the intermittency of large-scale coherent structures, which is accomplished using a computationally tractable continuous-discrete convolution approach applied to the inverse SPOD problem (Nekkanti & Schmidt 2021). Frequency−time diagrams of the expansion coefficients of $p$ were obtained via the convolution approach, as presented in Fig. 15. Overall, the high-energy regions across all the cases are predominantly concentrated within the low-frequency band of $St < 5$, aligning with the characteristic frequencies identified in the energy spectra of Fig. 13 and corresponding to the large-scale streaks and hairpin vortical events depicted in Fig. 14. However, compared with the low-frequency patches in the TNS (Fig. 15(a)) with some extent of temporal coherence, the low-frequency energy distribution in the CNS, as shown in Fig. 15(b), exhibits significantly fractured and fragmented patterns. Temporally, the high-energy contents diminish in scale and become increasingly scattered, along with an increased stratification between high-energy cores. These features suggest that the CNS attenuates the temporal continuity and smoothness of the turbulence structures observed in the TNS, thereby enhancing the turbulence's intermittency.

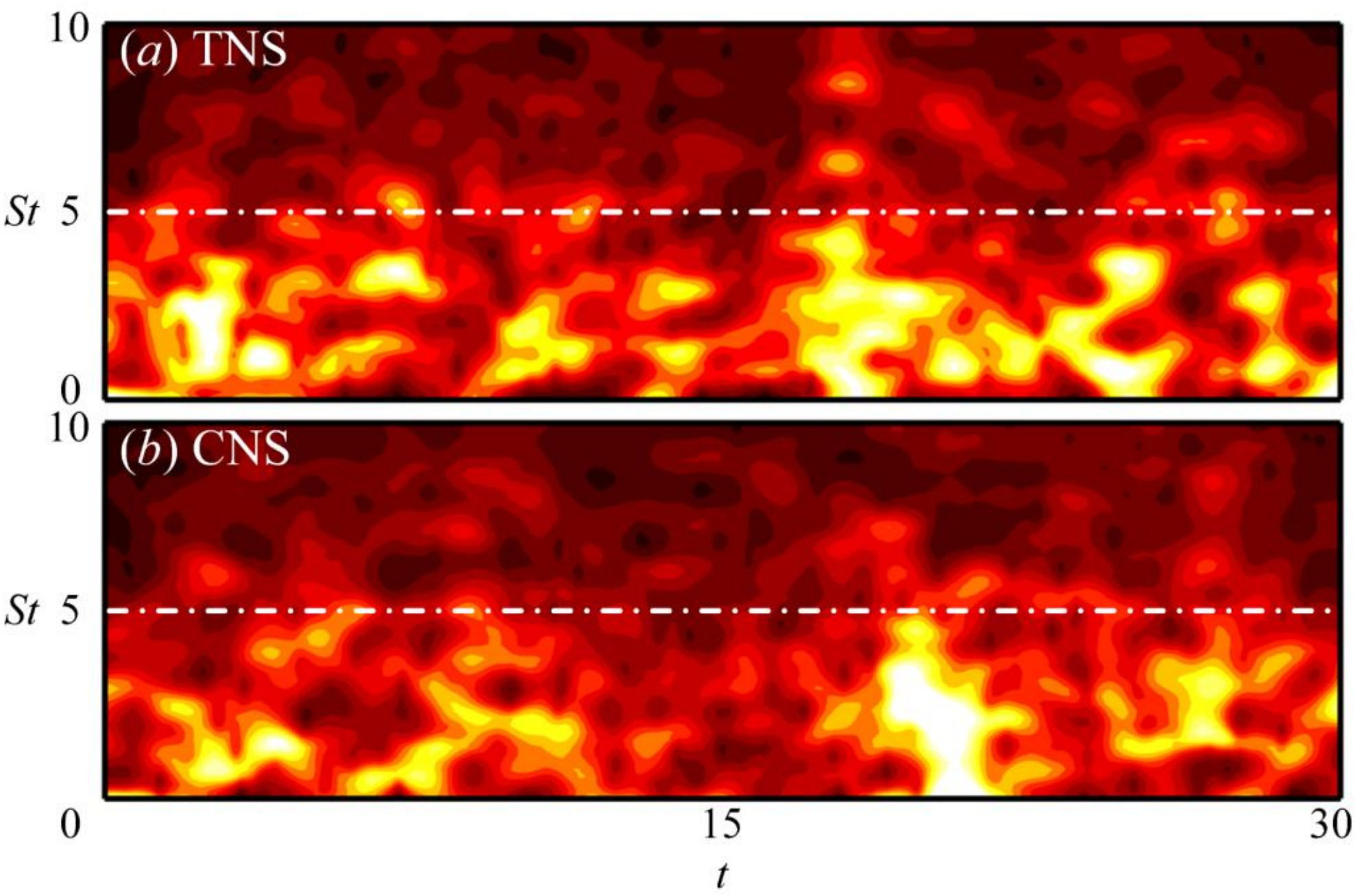

FIG. 15. SPOD-based frequency−time diagrams of $p$ for (a) TNS and (b) CNS models.

In summary, the application of SPOD provides a comprehensive spatiotemporal and frequency-domain investigations for evaluating the fundamental modifications introduced by the CNS relative to the TNS model. The SPOD approach unveils the convective nonlinearity phenomenon in the large-scale spatiotemporal structures of spectra-mode representations and confirms the scaling law relations of the −11/3 and −5/3 slopes for the pressure and Rortex in the small-scale turbulence, respectively. For the energy spectra, the CNS induces the characteristic frequency shifts that correspond to the longer streamwise wavelengths and denser vortex spacing. Spatiotemporally, the CNS model strengthens the wall-normal stratifications for the tilted vortex packets while degrading their overall coherence. Finally, the shearing-only mechanism enhances the structural intermittency by temporally fragmenting the high-energy vortex packets and shortening the dynamical lifecycles, meanwhile modulating the inter-scale energy cascade via the spatiotemporal harmonic phenomenon shifts.

## 5 Conclusions

The study interrogates the correction mechanism of the CNS equations via the DNS in fully developed turbulent channel flows at $Re_\tau = 550$ in order to validate the physical soundness of these equations. By systematically comparing the DNS data of CNS and TNS, the dynamic consequences are unveiled for the correction. The turbulence field is profoundly understood kinematically based on the RNRI enabled by Rortex. The SPOD approach brings out the convective nonlinearity phenomenon inherent in the CNS and TNS models, providing a unique insight into the spatiotemporal feature of the channel turbulence. These major findings and important conclusions are highlighted as following:

(1) From the perspective of fluid dynamics, the correction in CNS fundamentally adjusts the near-wall momentum transport in turbulence. Removal of the stretching viscous stresses induces an enhancement of near-wall velocity gradient, significantly thinning the viscous and buffer boundary layers. The mean streamwise velocity profile is then suppressed, which corrects the overshoot in TNS velocity and yields the excellent agreement with the experiment and DNS benchmarks. With the reference to these benchmarks, the current DNS data from CNS model does get improved when compared with the TNS model in terms of the mean velocity and the Reynolds stresses. However, the CNS model still needs to be further validated for its closeness to the fluid physics in reality when being applied in more accurate DNS algorithms, such as the code generating the LM data by Lee & Moser (2015).

(2) From the perspective of fluid kinematics, the current study identifies the strictly-defined vortex representation with the key elements uniquely determined by the VGT eigen status or discriminants ($1-2-3$R or $\Delta > 0, = 0, < 0$), namely the rotational/non-rotational interface (RNRI, $\Delta = 0$, 2D interface) as the vortex-cell boundary enwrapping a 3D vortical space, the rotational domain (RO, $\Delta > 0$, 3D iso-surface) and the Rortex core line (CL, $\nabla \boldsymbol{R} \times \boldsymbol{R} = 0$, 1D vector line) representing the vortex core center of local rotation axis, or the shortened form of vortex-cell structure (RNRI − RO − CL). Within the context, the CNS model is found to generate more compact and fragmented vortex-cells, such as the elongated quasi-streamwise vortices and distinct hairpin packets, through singling out the shearing-only mechanism in the governing equations.

(3) The analysis of the TKE budgets and wavenumber-premultiplied energy spectra elucidates the dynamical impact of the shearing-only constitutive relation. The CNS leads to a significant increase in both TKE production and isotropic dissipation, suggesting more vigorous vortex activity. The premultiplied one-dimensional energy spectra reveal a broadening and expansion of the high-energy core, accompanied by the elongation of near-wall coherent structures, indicating that the absence of stretching-induced viscous resistance facilitates the uninhibited growth of flow structures in the immediate vicinity of the wall.

(4) Regarding the characteristics of channel turbulence, the SPOD approach unveils the distinctive convection nonlinearity phenomenon for the DNS data in both CNS and TNS models, which well explains the mechanism of the spatiotemporal harmonics observed in the spectra and modal structures for the large-scale turbulence. The SPOD spectra further confirm the scaling law relations of $-5/3$ and $-11/3$ energy cascade slopes in the small-scale wall-bounded turbulence for Rortex and pressure, respectively. Moreover, the SPOD reveals that the CNS shearing-only mechanism alters the channel turbulence by modulating the energy cascade via frequency shifts, yielding the longer wavelengths and denser vortex spacings. Spatiotemporally, the strengthened shearing effects in CNS model induce more tilted vortex streaks in streamwise and more streak stratifications in wall-normal, as well as the enhanced intermittency by fragmenting vortex packets and degrading overall coherence.

Conclusively, with the fluid kinematics foundation being established on the VGT eigenvalues and discriminants, as represented by the Rortex, rotational/non-rotational interface (RNRI) and Rortex coreline (CL), the CNS model is expected to provide a new perspective on the intrinsic dynamics of wall-bounded turbulence. Within the context, the CNS, with its current evident physical soundness comparing to the TNS, needs to be applied to more DNS practices with high-order numerical accuracy and more challenging turbulent flows to better prove and further establish its physics validity and engineering applicability.

**Acknowledgements:** The authors acknowledge that the DNS data in channel was publicly provided by Myoungkyu Lee and Robert D. Moser in the Department of Mechanical Engineering of the University of Texas at Austin and the SPOD source code was obtained from Oliver T. Schmidt in the Department of Mechanical and Aerospace Engineering at University of California San Diego.

**Funding statement:** The research was funded by the Shanghai Municipal Commission of Education (Grant No. AR960).

**Competing interests:** The authors declare that there are not any competing financial interests or personal relationships that could have appeared to influence the work reported in the paper.

**Data availability statement:** Data are available by contacting the corresponding author.

**Author contributions:** **Y.W.:** Conceptualization, Data curation, Formal analysis, Investigation, Methodology, Software, Visualization, Writing − original draft. **D.W.:** Methodology, Software, Writing − review & editing. **S.Z.:** Methodology, Software, Validation. **H.X.:** Conceptualization, Funding acquisition, Methodology, Supervision, Writing − review & editing.